\documentclass{amsart}
\usepackage{graphicx}
\usepackage{amssymb}
\usepackage{lscape}
\usepackage[hyperindex]{hyperref}
\usepackage{epstopdf}
\usepackage{amssymb,amsmath}
\usepackage{subfigure}
\usepackage{setspace}
\usepackage[normalem]{ulem}
\usepackage{verbatim}
\usepackage{color}
\usepackage{anyfontsize}
\usepackage{wrapfig}
\usepackage{sidecap, enumerate}

\usepackage{natbib}

\usepackage{xr}
\newtheorem{theorem}{Theorem}[section]

\newenvironment{remark}[1][Remark]{\begin{trivlist}
\item[\hskip \labelsep {\bfseries #1}]}{\end{trivlist}}

\begin{document}

\title[A Nonparametric Bayesian Technique for High-Dimensional Regression]{A Nonparametric Bayesian Technique for High-Dimensional Regression}

\author[Guha]{Subharup Guha}
\address{University of Missouri, USA.}
\email{GuhaSu@missouri.edu}
\author[Baladandayuthapani]{Veerabhadran Baladandayuthapani}
\address{The University of Texas MD Anderson Cancer Center and Rice University, USA.}
\email{Veera@mdanderson.org}
\thanks{Subharup Guha is Associate Professor, Department of Statistics, University of Missouri; Veerabhadran Baladandayuthapani is Associate Professor, Department of Biostatistics, The University of Texas MD Anderson Cancer Center, and Adjunct Associate Professor, Department of Biostatistics, Rice University (email addresses: GuhaSu@missouri.edu, Veera@mdanderson.org).  This work
was supported by the National Science Foundation under awards DMS-0906734, DMS-1461948
 to SG and DMS-1463233 to VB, and by the National Institutes of Health under grant R01 CA160736 to VB. The authors thank Ganiraju Manyam for help with biological interpretations of the results and Chiyu Gu for key ideas about the Markov chain Monte Carlo procedure. }

\begin{abstract}

This paper proposes a nonparametric Bayesian framework called \textbf{VariScan} for simultaneous clustering, variable selection, and prediction in high-throughput  regression settings. Poisson-Dirichlet processes are utilized to detect lower-dimensional latent clusters of covariates. An adaptive nonlinear prediction model is constructed for the response, achieving a balance between model parsimony and~flexibility.
Contrary to conventional belief, cluster detection is shown to be aposteriori consistent for a general class of models as the number of covariates and subjects grows. Simulation studies and data analyses demonstrate that VariScan often outperforms several well-known statistical methods.

\end{abstract}

 \maketitle

\section{Introduction}\label{S:introduction}

An increasing number of studies
 involve the regression analysis of $p$ continuous covariates and continuous or discrete univariate responses on $n$ subjects, with $p$ being much larger than $n$.
The development of effective clustering and sparse regression models for reliable predictions is especially challenging in these ``small $n$, large $p$''  problems.
The goal of the analysis is often three-pronged: {\it{(i) Cluster identification:}} We wish to identify clusters of covariates with similar patterns for the subjects. For example,  in biomedical studies where  the covariates are gene expression levels,   subsets of  genes associated with distinctive between-subject patterns may correspond to different underlying biological processes; {\it{(ii) Detection of sparse regression predictors:}} From the set of $p$ covariates, we wish to select a sparse subset of reliable predictors for the subject-specific responses and infer the nature of their relationship with the responses.  In most genomic applications, just a few of the biological processes  are usually  relevant to a response variable of interest, and we need reliable and parsimonious regression models; and {\it{(iii) Response prediction:}} Using the inferred regression relationship, we wish to  predict the responses of $\tilde{n}$ additional subjects for whom only covariate information is available. The reliability of an inference procedure is measured by its prediction accuracy for out-of-sample  individuals.

In high-throughput  regression settings  with continuous covariates and  continuous or discrete outcomes, this paper proposes a nonparametric Bayesian framework called \textbf{VariScan} for simultaneous clustering, variable selection, and prediction.
\subsection{Motivating applications}

Our methods and computational endeavors are motivated by recent high-throughput investigations in biomedical research, especially in cancer. Advances in array-based technologies allow simultaneous measurements of biological units (e.g.\ genes) on a relatively small number of subjects. Practitioners wish to select important genes involved with the disease processes and to develop efficient prediction models for patient-specific clinical outcomes such as continuous survival times or categorical disease subtypes. The analytical challenges posed by such data include not only high-dimensionality but also the existence of considerable gene-gene correlations induced by biological interactions. In this article, we analyze gene expression profiles assessed using microarrays in patients with diffuse large B-cell lymphoma (DLBCL)  \citep{Rosenwald_etal_2002} and breast cancer  \citep{vantVeer_2002}. Both datasets are publicly available and  have the following general characteristics: for individuals $i=1,\ldots,n$, the data consist of mRNA expression levels $x_{i1},\ldots,x_{ip}$ for $p$ genes, where $n>>p$, along with censored survival times denoted by $w_i$. More details, analysis results, and gains using our methods over competing approaches are  discussed in Section~\ref{S:benchmark_data}.

The scope and success of the proposed methodology and its associated theoretical results extend far beyond the  examples we  discuss in this paper. For instance, the technique is not restricted to biomedical studies; we have successfully applied VariScan in a variety of other high-dimensional~applications and achieved high inferential gains relative to existing~methodologies.

\subsection{Challenges in high-dimensional predictor detection}
Despite the large number of existing methods related to clustering, variable selection and prediction, researchers continue to develop new methods to meet the challenges posed by newer applications and larger datasets. Predictor detection becomes particularly problematic in big datasets due to the pervasive collinearity of the covariates.

\begin{figure}
\begin{center}
\includegraphics[scale=0.25]{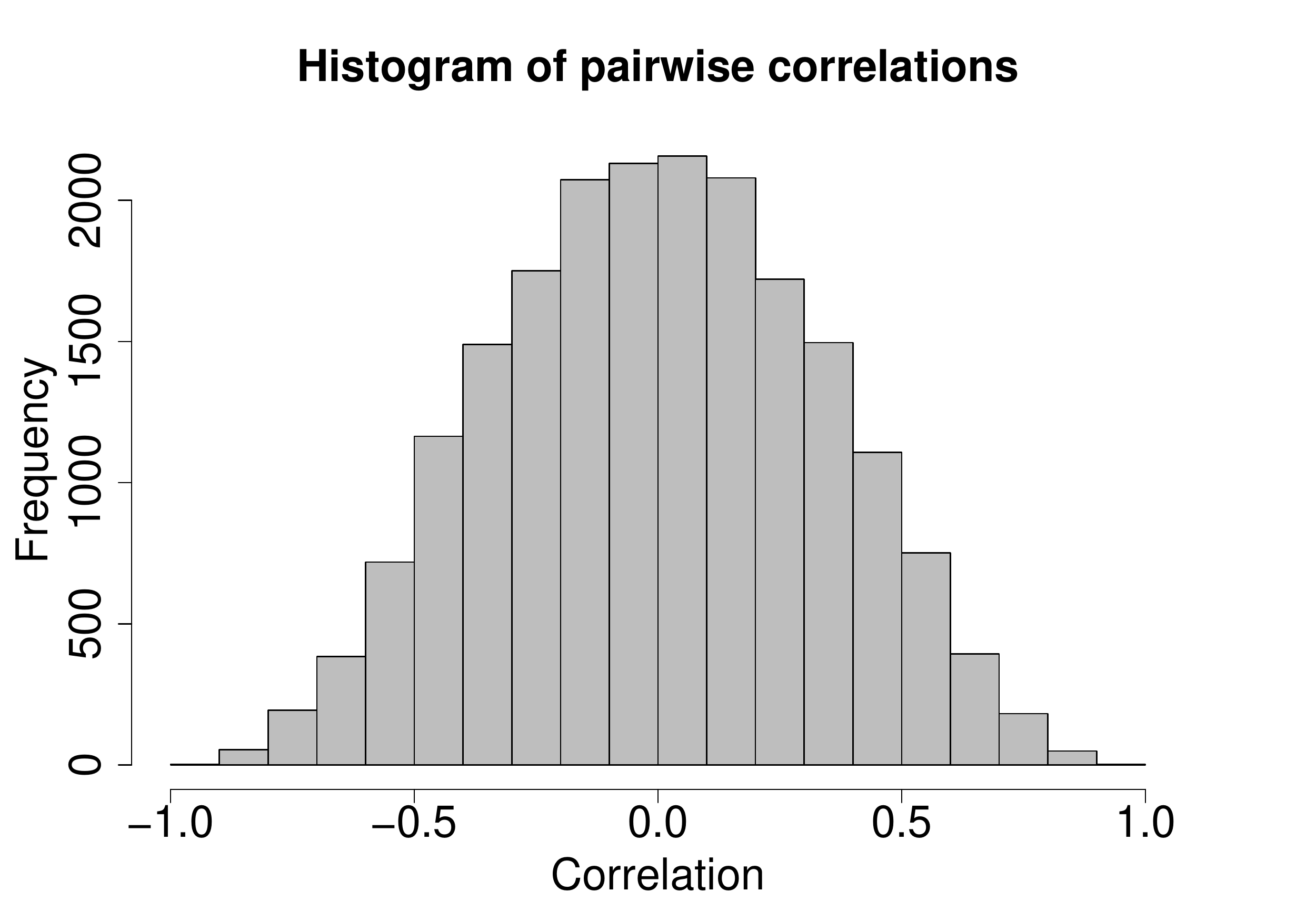}
\caption{Pairwise sample correlations for $p=200$ vectors independently generated from a multivariate normal distribution with $n=10$ uncorrelated elements.}
\label{F:toy_pair}
\end{center}
\end{figure}

For a simple demonstration of this fact, consider a process that independently samples  $n$-variate covariate column vectors $\boldsymbol{x}_1\ldots,\boldsymbol{x}_p$, so that $p=200$  vectors with $n=10$ i.i.d.\ elements are generated from a common normal distribution.   Although the vectors are independently generated, extreme values of the pairwise correlations are  observed in the sample, as shown in the histogram of Figure \ref{F:toy_pair}. The proportion of extremely high or low correlations typically increases with $p$, and with greater correlation of the generated vectors under the true~process.

Multicollinearity is common in high-dimensional problems because the $n$ - dimensional space of the covariate columns becomes saturated with the large number of covariates. This is disadvantageous for regression because a cohort of highly correlated covariates is weakly identifiable as  regression predictors.
For {example,
 imagine that the $j^{th}$ and $k^{th}$ covariate columns have a sample correlation close to 1, but that neither covariate is really a predictor in a linear regression model. An  alternative  model in which  \textit{both} covariates are predictors  with equal and opposite regression coefficients, has a nearly  identical joint likelihood for all regression outcomes. Consequently, an inference procedure is often unable to choose between these competing models as the likely explanation for the data.

 In the absence of  strong application-specific priors to guide  model selection, collinearity  makes it  impossible to pick the true set of predictors in high-dimensional problems.
 Furthermore, collinearity  causes unstable  inferences  and  erroneous test case predictions \citep{Weisberg_1985}. The problem is  exacerbated if some of the regression outcomes are unobserved, as with categorical responses and survival~applications.

\subsection{Bidirectional clustering with adaptively nonlinear functional regression and prediction}

 Since the data in small $n$, large $p$ regression problems  are informative only about  the combined effect  of a cohort of highly correlated covariates, we address the  issue of collinearity using clustering approaches. Specifically, VariScan utilizes the sparsity-inducing property of Poisson-Dirichlet processes (PDPs) to first group the $p$   columns of the covariate matrix  into $q$  latent clusters, where $q \ll p$, with each cluster consisting of columns with similar patterns across the subjects.   The data are allowed to direct the choice  between a class of PDPs and their special case, a Dirichlet process, for selecting a suitable allocation scheme for the covariates. These partitions could provide meaningful insight into unknown biological processes (e.g.\ signaling pathways) represented by the latent clusters.

 To flexibly capture the within-cluster pattern  of the covariates, the $n$ subjects are allowed to group differently in each cluster via a nested Dirichlet process. This feature is motivated by genomic studies \citep[e.g.,][]{Jiang_Tang_Zhang_2004} which have demonstrated that subjects or biological samples often group differently under different biological processes. In essence, this modeling framework specifies a random, bidirectional (covariate, subject) nested clustering of the high-dimensional covariate~matrix.

Clustering downsizes the small $n$, large $p$  problem to a ``small $n$, small $q$'' problem, facilitating an effective stochastic  search of the indices $\mathcal{S}^* \subset\{1,\ldots,q\}$ of potential \textit{cluster predictors}. If necessary, we could then infer the indices $\mathcal{S}\subset\{1,\ldots,p\}$ of the  covariate predictors. This feature differentiates the VariScan procedure from  black-box nonlinear prediction methods. In addition, the technique is capable of detecting functional relationships through elements such as nonlinear functional kernels and basis functions such as splines or wavelets. An adaptive mixture of linear and nonlinear elements in the regression relationship aims to achieve a balance between model parsimony and~flexibility. These aspects of VariScan define a joint model for the responses and covariates, resulting in an effective model-based clustering and variable selection procedure, improved posterior
inferences and accurate test case~predictions.

\begin{figure}
\begin{center}
\includegraphics[scale=0.6]{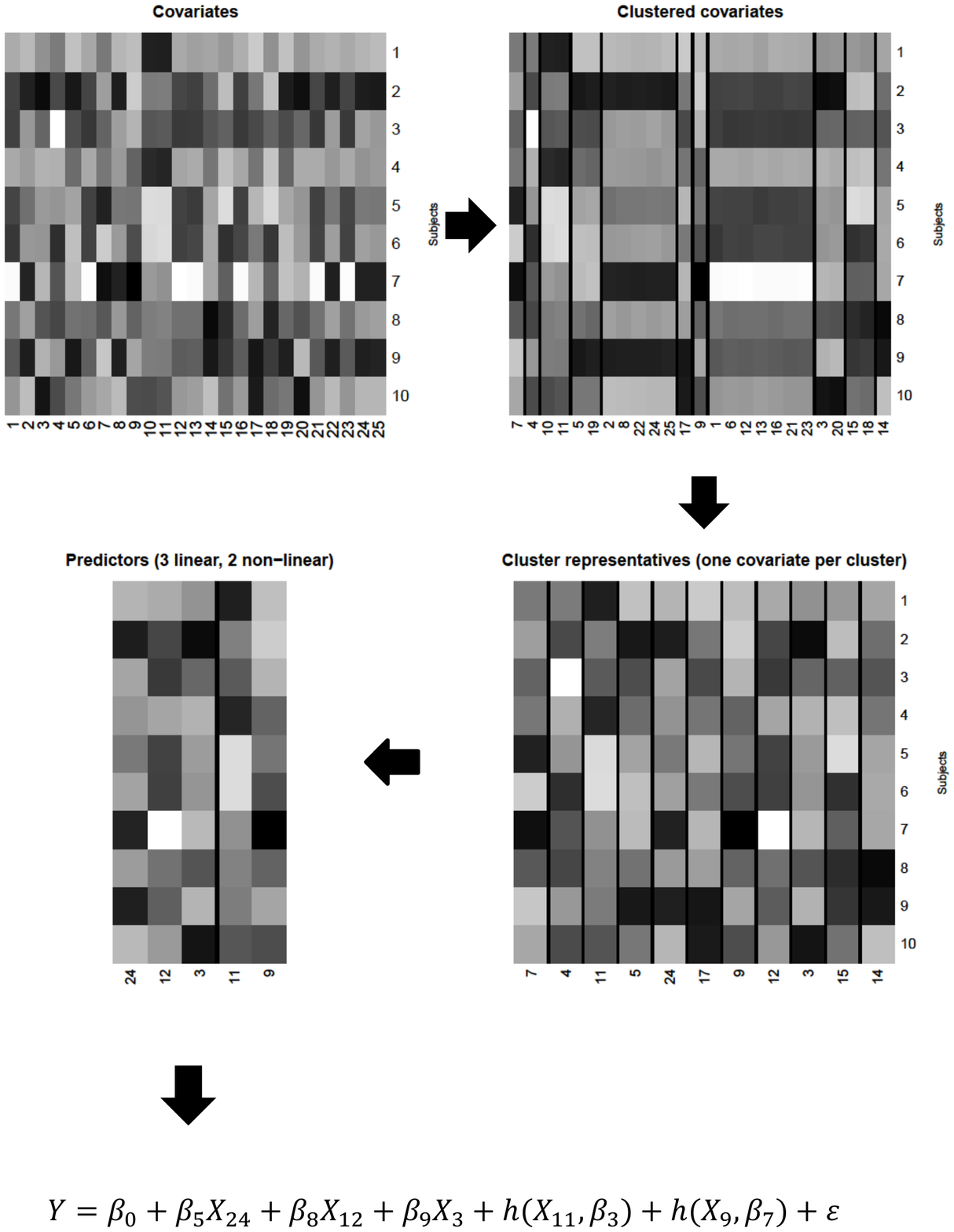}
\vspace{-2.5 cm}
\caption{Stylized example illustrating the basic methodology for reliable prediction for $n=10$ subjects and $p=25$ covariates allocated to $q=11$ number of PDP clusters. The column labels represent the covariate indices. The row labels are the subjects. See the text for further explanation.}
\label{F:toy_VariScan}
\end{center}
\end{figure}

Figure \ref{F:toy_VariScan}  illustrates the key ideas of VariScan  using a toy example with $n=10$ subjects and $p=25$ covariates. The plot in the upper left panel represents a heatmap of the covariates. When investigators are interested in discovering a sparse prediction model for additional subjects, the posterior analysis averages over all possible realizations of two basic steps, both of which  are stochastic and may be stylistically described as follows:

 \begin{enumerate}
 \item\textbf{Clustering} \quad The column vectors are allocated in an unsupervised manner to $q=11$ number of PDP clusters. This is plotted in the upper right panel, where the columns are grouped via bidirectional clustering to reveal the similarities in the within-cluster~patterns.

     \item\textbf{Variable selection and regression} \quad One covariate is stochastically selected from each cluster and is known as the \textit{cluster representative}. The middle right panel displays the set of representatives, $\{\boldsymbol{x}_7,\boldsymbol{x}_4,\boldsymbol{x}_{11},\boldsymbol{x}_5,\boldsymbol{x}_{24},$
$\boldsymbol{x}_{17},\boldsymbol{x}_9,\boldsymbol{x}_{12},\boldsymbol{x}_3,
\boldsymbol{x}_{15},\boldsymbol{x}_{14}\}$, for the 11 clusters. The regression predictors are stochastically selected from the random set of the cluster representatives. Some  representatives are not associated with the response; the remaining covariates are outcome  predictors and may    have either a linear or nonlinear regression relationship. The linear predictors  $\{\boldsymbol{x}_{24},\boldsymbol{x}_{12},\boldsymbol{x}_{3}\}$ and non-linear predictors $\{\boldsymbol{x}_{11},\boldsymbol{x}_{9}\}$ are shown in the middle left panel. For a nonlinear function $h$, the regression equation for a subject is displayed in the lower panel for a zero-mean Gaussian error, $\epsilon$. The subscripts of the $\beta$ parameters are the cluster labels, e.g., covariate $\boldsymbol{x}_{24}$ represents the fifth PDP cluster.
\end{enumerate}
When out-of-the-bag prediction is not of primary interest, alternative variable selection strategies  discussed in Section \ref{S:predictors} may be applied.

\subsection{Existing Bayesian approaches and limitations}\label{S:lit review}

There is a vast literature on  Bayesian strategies for one or more of the three inferential goals mentioned at the beginning of Section \ref{S:introduction}. A majority of Bayesian model-based clustering techniques rely on the celebrated Dirichlet process; see \citet[chap.\ 4]{muller2013bayesian} for a comprehensive review. A seminal paper by \cite*{Lijoi_Mena_Prunster_2007b} advocated the use of Gibbs-type priors \citep*{Gnedin_Pitman_2005, Lijoi_Mena_Prunster_2007a} for accommodating more flexible clustering mechanisms than those induced by the Dirichlet process. This work also demonstrated the  practical utility of PDPs in genomic applications.

Among model-based clustering techniques based on Dirichlet processes, the approaches of \cite{Medvedovic_Siva_2002}, \cite{Dahl_2006}, and \cite{Muller_Quintana_Rosner_2011} assume that it is possible to \textit{globally} reshuffle the rows and columns of the covariate matrix to reveal the clustering pattern. More closely related to our clustering  objectives is  the nonparametric Bayesian local clustering (NoB-LoC) approach of \cite{Lee_etal_2013}, which   clusters  the covariates \textit{locally} using two sets of Dirichlet processes. Although some similarities exist between NoB-LoC and the clustering aspect of VariScan, there are major differences. Specifically, the  VariScan framework can accommodate high-dimensional regression in addition to bidirectional clustering. Furthermore, VariScan typically produces more efficient inferences by its greater flexibility in modeling a larger class of clustering patterns via PDPs. The distinction becomes especially important for  genomic datasets  where  PDP-based models are often preferred to Dirichlet-based models by log-Bayes factors on the order of thousands; see Section \ref{S:benchmark_data} for an example. Moreover, the Markov chain Monte Carlo (MCMC) implementation of VariScan explores the posterior substantially faster due to its better ability to  allocate outlying covariates to singleton clusters via augmented variable Gibbs sampling. From a theoretical perspective, contrary to widely held beliefs about the  non-identifiability of mixture model clusters, we discover the remarkable theoretical property of VariScan that, as both $n$ and $p$  grow, a fixed set of  covariates that (do not) co-cluster under the true VariScan model,  also (do not) asymptotically co-cluster under its posterior.}

{From a regression-based Bayesian viewpoint,  perhaps the most ubiquitous approaches are based on Bayesian variable selection techniques  in linear and non-linear regression models. See \cite{david2002bayesian} for a comprehensive review.  For Gaussian responses, the common linear methods include stochastic search variable selection \citep{George_McCulloch_1993}, selection-based priors \citep{Kuo_Mallick_1997} and shrinkage-based methods \citep{Park_Casella_2008, xu2015bayesian, griffin2010inference}.  Some of these methods have been extended to non-linear regression contexts \citep{smith1996nonparametric} and  to  generalized linear models \citep{dey2000generalized,meyer2002predictive}. However, most of the {afore-mentioned regression} methods are based on {strong} parametric assumptions   and do not explicitly account for the multicollinearity commonly observed in high-dimensional settings. Nonparametric approaches typically assume priors on the error residuals \citep{hanson2002modeling,kundu2014bayes} or on the regression coefficients using random effect representations \citep{bush1996semiparametric, maclehose2010bayesian}. For nonparametric mean function estimations, they are typically based on basis function expansions such as wavelets\citep{morris2006wavelet} and splines \citep{baladandayuthapani2005spatially}. We take a fundamentally different approach in this article by defining a nonparametric joint model, first  on the covariates and then via an adaptive nonlinear prediction model on the responses.}

The rest of the paper is organized as follows. We introduce the VariScan model in Section \ref{S:VariScan_model}. 
Some theoretical results for the VariScan procedure are presented in Sections \ref{S:clusters}. Through simulations in Section \ref{S:simulation2} and \ref{S:simulation}, we demonstrate the accuracy of the clustering mechanism and compare the prediction reliability of VariScan  with that of several  established variable selection procedures using artificial datasets. In Section \ref{S:benchmark_data}, we analyze the motivating  gene expression microarray
datasets for leukemia and breast cancer to demonstrate the  effectiveness of VariScan  and compare its prediction accuracy with those of competing methods. Additional supplementary materials contain the theorem proofs, as well as additional simulation and  data analyses~results.

\bigskip


\section{VariScan Model}\label{S:VariScan_model}

In this section, we layout  the detailed construction of the Variscan model components, which involves two major steps. First, we utilize the sparsity-inducing property of Poisson-Dirichlet processes to perform a directional nested clustering of the covariate matrix (Section \ref{S:covariates}), and second, we describe the choice of the cluster-specific predictors  and nonlinearly relate them to Gaussian regression outcomes of the subjects (Section \ref{S:predictors}). Subsequently, Section \ref{S:justifications}  provides details of the  model justifications and generalizations to discrete and survival outcomes.

\subsection{Covariate clustering model}\label{S:covariates}

First, each  of the $p$ covariate matrix columns, $\boldsymbol{x}_{1},\ldots,\boldsymbol{x}_{p}$, is assigned to one of $q$ latent  clusters, where $q \ll p$, and where the assignments and  $q$ are  unknown.
That is, for $j=1,\ldots,p$ and $k=1,\ldots,q$, an \textbf{allocation variable}
$c_j$    equals  $k$ if the $j^{th}$ covariate is assigned to the $k^{th}$ cluster.

We posit that the $q$ clusters are associated with  \textbf{latent vectors} $\boldsymbol{v}_1,\ldots,\boldsymbol{v}_q$  of length $n$. The covariate columns assigned to a latent cluster are essentially contaminated versions of these cluster's latent vector and thus induces  high correlations among covariates belonging to a cluster.
In practice, however, a few individuals within each  cluster may have highly variable covariates. We model this aspect by associating a larger error variance with those individuals. This is achieved via a Bernoulli variable,~$z_{ik}$, for which the value $z_{ik}=0$ indicates a high variance:
\begin{align*}
x_{ij} \mid z_{i  k}, c_j = k  &\stackrel{indep}\sim
    \begin{cases}
       N(v_{ik}, \tau_1^2) \quad &\text{if  $z_{ik}=0$}\\
         N(v_{ik}, \tau^2) \quad&\text{if  $z_{ik}=1$}\\
    \end{cases}
\end{align*}
 where
$\tau_1^2$ and $\tau^2$ are variance parameters with inverse Gamma priors and $\tau_1^2$ is much greater than $\tau^2$. It is assumed that the support of $\tau$ is bounded below by a small, positive constant, $\tau_*$. Although not necessary from a methodological perspective, this restriction guarantees the asymptotic result of Section \ref{S:clusters}.
The indicator variables for the individual--cluster combinations are  apriori modeled {\it i.i.d} as:
\begin{equation*}
z_{ik} \stackrel{iid}\sim \text{Ber}(\xi),  \qquad\text{$i=1,\ldots,n$ and $k=1,\ldots,q$,}
\end{equation*}
where $\xi \sim \text{beta}(\iota_1,\iota_0)$. The condition  $\iota_1 \gg \iota_0$ guarantees that  prior probability $P(z_{ik} = 1)$ is nearly equal to 1, and so only a small proportion of the individuals have highly variable covariates within each cluster.

\bigskip

\noindent \textbf{Allocation variables.} As an appropriate model for the covariate-to-cluster allocations that accommodates a wide range of allocation patterns, we rely on the partitions induced by the \textit{two-parameter Poisson-Dirichlet process}, $\mathbb{PDP}\bigl(\alpha_1, d\bigr)$, with discount parameter $0 \le d < 1$ and precision or mass parameter $\alpha_1>0$. In genomic applications, for example, these partitions may allow the discovery of unknown biological processes represented by the latent clusters.
  We defer additional details of the empirical and theoretical justifications of using PDP processes until Section~\ref{S:justifications}.

The PDP was introduced by \cite{Perman_etal_1992} and later investigated by \cite{Pitman_1995} and \cite{Pitman_Yor_1997}. Refer to \cite{Lijoi_Prunster_2010} for a detailed discussion of different classes of Bayesian nonparametric~models, including Gibbs-type priors \citep*{Gnedin_Pitman_2005, Lijoi_Mena_Prunster_2007a} such as Dirichlet processes and PDPs. \cite*{Lijoi_Mena_Prunster_2007b} were the first to implement Gibbs-type priors for more flexible clustering mechanisms than  Dirichlet process partitions.

The  PDP-based allocation variables are apriori exchangeable and  evolve as follows. Since the cluster allocations labels are arbitrary, we may assume without loss of generality that  $c_1=1$, i.e., the first covariate is assigned to the first cluster.  Subsequently, for covariates $j=2,\ldots,p$, suppose there exist
   $q^{(j-1)}$ distinct clusters among $c_1,\ldots,c_{j-1}$, with the $k^{th}$ cluster containing  $n_{k}^{(j-1)}$ number of covariates. The  predictive probability that the $j^{th}$ covariate is assigned to the $k^{th}$ cluster is~then
\begin{align*}
P\left(c_j = k \mid c_1, \ldots, c_{j-1} \right) \propto
    \begin{cases}
        n_{k}^{(j-1)} - d 
         \quad &\text{if $k = 1,\ldots,q^{(j-1)}$}\\
        \alpha_1 + q^{(j-1)} \cdot d \quad &\text{if $k = q^{(j-1)} + 1$}\\
        \end{cases}
\end{align*}
where the event $c_j=q^{(j-1)} + 1$ in the second line corresponds to the $j^{th}$ covariate opening a new cluster.
 When $d = 0$, we obtain the well known P\`{o}lya urn scheme for Dirichlet processes \citep{Ferguson_1973}.

 In general, exchangeability holds for all product partition models \citep{Barry_Hartigan_1993, Quintana_Iglesias_2003} and species sampling models \citep{Ishwaran_James_2003}, of which PDPs are a special case. The number of  clusters, $q$,  stochastically increases as $\alpha_1$ and $d$ increase. For  $d$ fixed,  the $p$ covariates are each assigned to $p$ singleton clusters in the limit as $\alpha_1 \to \infty$. 

 A PDP achieves dimension reduction in the number of covariates because $q$, the random number of clusters, is asymptotically equivalent to
\begin{align}
    \begin{cases}
        \alpha_1 \cdot \log p        \qquad &\text{if $d = 0$} \\ 
        T_{d, \alpha_1} \cdot p^d\qquad &\text{if $0 < d < 1$}\\
        \end{cases}\label{q}
\end{align}
for  a random variable $T_{d, \alpha_1} >0$ as $p\to \infty$. This implies that the number of  Dirichlet process clusters (i.e., when $d=0$) is asymptotically of a smaller order than the number of   PDP  clusters when $d>0$. This property was effectively utilized by \cite*{Lijoi_Mena_Prunster_2007b} in species prediction problems and applied to gene discovery settings. The use of Dirichlet processes to achieve
 dimension reduction has precedence in the literature; see \cite{Medvedovic_etal_2004}, \cite{Kim_etal_2006}, \cite{Dunson_etal_2008} and \cite{Dunson_Park_2008}.

The PDP discount parameter $d$ is given the mixture prior $\frac{1}{2}\delta_0 + \frac{1}{2} U(0,1)$, where $\delta_0$ denotes the point mass at 0. Posterior inferences of this parameter allows us to flexibly choose between Dirichlet processes and more general~PDPs for the best-fitting clustering mechanism.

\bigskip

\noindent \textbf{Latent vectors.} The hierarchical prior for the covariates is completed by specifying a \textit{base  distribution} $G^{(n)}$ in $\mathcal{R}^n$ for the latent vectors $\boldsymbol{v}_1,\ldots,\boldsymbol{v}_q$. Consistent with our goal of developing a flexible and scalable inference procedure capable of fitting large datasets, we impose additional lower-dimensional structure on the $n$-variate base distribution. Specifically, since the $n$ subjects are exchangeable, base distribution $G^{(n)}$ is assumed to be the $n$-fold product measure of a univariate distribution, $G$. This allows the individuals and clusters to communicate through the $nq$ number of   latent vector elements:
\begin{equation}
v_{ik} \stackrel{iid}\sim G \qquad \text{for } i=1,\ldots,n, \text{ and } k=1,\ldots,q. \label{v}
\end{equation}
The unknown, univariate distribution, $G$,  is itself given  a nonparametric Dirichlet process prior,  allowing the latent vectors to flexibly capture the within-covariate patterns of the subjects:
\begin{equation}
G \sim \mathcal{DP}(\alpha_2)\label{G}
\end{equation}
for   mass parameter $\alpha_2>0$ and univariate base distribution,  $N(\mu_2, \tau_2^2)$. Being a realization of a Dirichlet process, distribution $G$ is discrete and allows the subjects to group differently in different PDP clusters.  
The  number of distinct values among the $v_{ik}$'s  is  asymptotically equivalent to $\alpha_2 \cdot \log n  q$, facilitating further dimension reduction and scalability of inference as $n$ approaches hundreds or thousands of individuals, as commonly encountered  in genomic datasets.

\smallskip

\subsection{Prediction and regression model}\label{S:predictors}

Now, suppose there are $n_{k}$ covariates  allocated to the  $k^{th}$ cluster. 
We posit that each cluster elects from among its   member covariates a  \textit{representative}, denoted by $\boldsymbol{u}_k$. A subset of the $q$ cluster representatives, rather than the covariates, feature in an additive regression model that can  accommodate nonlinear functional relationships.  The cluster representatives may be chosen in several different ways depending on the  application. Some possible options include:
\begin{enumerate}
\item[\textit{(i)}] Select with apriori equal probability one of the $n_k$  covariates belonging to the $k^{th}$ cluster  as the representative. Let $s_k$ denote the index of the covariate chosen as the representative, so that $c_{s_k}=k$ and $\boldsymbol{u}_k=\boldsymbol{x}_{s_k}$.
\item[\textit{(ii)}]     Set  latent vector $\boldsymbol{v}_k$ of Section \ref{S:covariates} as the cluster representative.
\end{enumerate}
Option \textit{(i)} is preferable when practitioners are mainly interested in identifying the effects of individual regressors, as in gene selection applications in cancer survival times (as noted in the introduction). Option \textit{(ii)} is  preferable when the emphasis is less on covariate selection and more on identifying clusters of candidate variables (e.g., genomic pathways) that are jointly associated with the responses.

The regression predictors are selected from among the $q$ cluster representatives, with their  parent clusters constituting the set of \textit{cluster predictors}, $\mathcal{S}^* \subset\{1,\ldots,q\}$.
Extensions of the spike-and-slab approaches \citep{George_McCulloch_1993, Kuo_Mallick_1997, Brown_etal_1998} are applied to relate the regression outcomes to the cluster representatives as:
\begin{align}
y_i &\stackrel{indep}\sim N\left( \eta_i,\, \sigma_i^2\right), \quad\text{where} \notag\\
\eta_i &= \beta_0 + \sum_{k=1}^q \gamma_{k}^{(1)}  \beta_k^{(1)} u_{ik} + \sum_{k=1}^q \gamma_{k}^{(2)}  h(u_{ik},\boldsymbol{\beta}_k^{(2)})  \label{eta_i}
\end{align}
and $h$  is  a nonlinear function.  Possible options for the nonlinear function $h$ in equation (\ref{eta_i}) include reproducible kernel Hilbert spaces \citep{mallickJRSSB2005}, nonlinear basis smoothing splines \citep{Eubank1999}, and wavelets. Alternatively, due to their interpretability as a linear model,   order-$r$ splines with $m$ number of knots \citep{deBoor_1978, Hastie_Tibshirani_1990,Denison_etal_1998} are especially attractive and computationally tractable.

The linear predictor $\eta_i$ in expression (\ref{eta_i}) implicitly relies on a vector of cluster-specific indicators, $\boldsymbol{\gamma}=(\boldsymbol{\gamma}_1,\ldots,\boldsymbol{\gamma}_q)$, where the triplet of indicators, $\boldsymbol{\gamma}_k=(\gamma_{k}^{(0)}, \gamma_{k}^{(1)}, \gamma_{k}^{(2)})$, add to 1 for each cluster $k$. If $\gamma_{k}^{(0)}=1$, the cluster representative and none of the covariates belonging to cluster~$k$ are associated with the responses.  If $\gamma_{k}^{(1)}=1$, the cluster representative appears as a simple linear regressor  in equation (\ref{eta_i});  $\gamma_{k}^{(2)}=1$ implies a nonlinear regressor.  
The number of linear predictors, non-linear predictors, and non-predictors are respectively, $q_1 =\sum_{j=1}^q \gamma_j^{(1)}$,  $q_2 =\sum_{j=1}^q \gamma_j^{(2)}$, and $q_0 =q-q_1 - q_2$.
For a simple illustration of this concept, consider again the toy example of Figure \ref{F:toy_VariScan}, where  one covariate is nominated from each cluster as the~representative. Of the $q=11$ cluster representatives, $q_1=3$ are linear predictors, $q_2=2$ are non-linear predictors, and the remaining $q_0=6$ representatives are non-predictors.

For nonlinear functions $h$ having a linear representation (e.g., splines), let $\boldsymbol{U}_{\boldsymbol{\gamma}}$ be  a  matrix of $n$ rows  consisting of the intercept column and the independent regressors based on the cluster representatives. For example, if we use order-$r$ splines with $m$ number of knots in equation (\ref{eta_i}), then the number of columns, $\text{col}(\boldsymbol{U}_{\boldsymbol{\gamma}})=q_1 + (m+r)\cdot q_2 + 1$. With $[\cdot]$ denoting  densities of random variables, the
 prior,
\begin{equation}
[\boldsymbol{\gamma}] \propto \omega_0^{q_0}\omega_1^{q_1}\omega_2^{q_2}\cdot \mathcal{I}\biggl(\text{col}(\boldsymbol{U}_{\boldsymbol{\gamma}}) <   n\biggr), \label{gamma}
\end{equation}
where the probabilities $\boldsymbol{\omega}=(\omega_{0}, \omega_{1}, \omega_{2})$ are given the Dirichlet distribution prior, $\boldsymbol{\omega} \sim \mathcal{D}_3(1,1,1)$.
The truncated prior for $\boldsymbol{\gamma}$ is designed to ensure model sparsity and prevent overfitting, as explained below.  Conditional on the variances of the regression outcomes in equation (\ref{eta_i}), we postulate a weighted g~prior for the regression coefficients:
\begin{equation}
\boldsymbol{\beta}_{\boldsymbol{\gamma}} | \boldsymbol{\Sigma} \sim N_{|\mathcal{S}^*|+1}\biggl(\boldsymbol{0}, \sigma_\beta^2({\boldsymbol{U}_{\boldsymbol{\gamma}}}'\boldsymbol{\Sigma}^{-1}\boldsymbol{U}_{\boldsymbol{\gamma}})^{-1}\biggr),\label{Zellner}
\end{equation}
where matrix $\boldsymbol{\Sigma}=\text{diag}(\sigma_1^2,\ldots,\sigma_n^2)$.

\begin{figure}
\begin{center}
\includegraphics[scale=1.1]{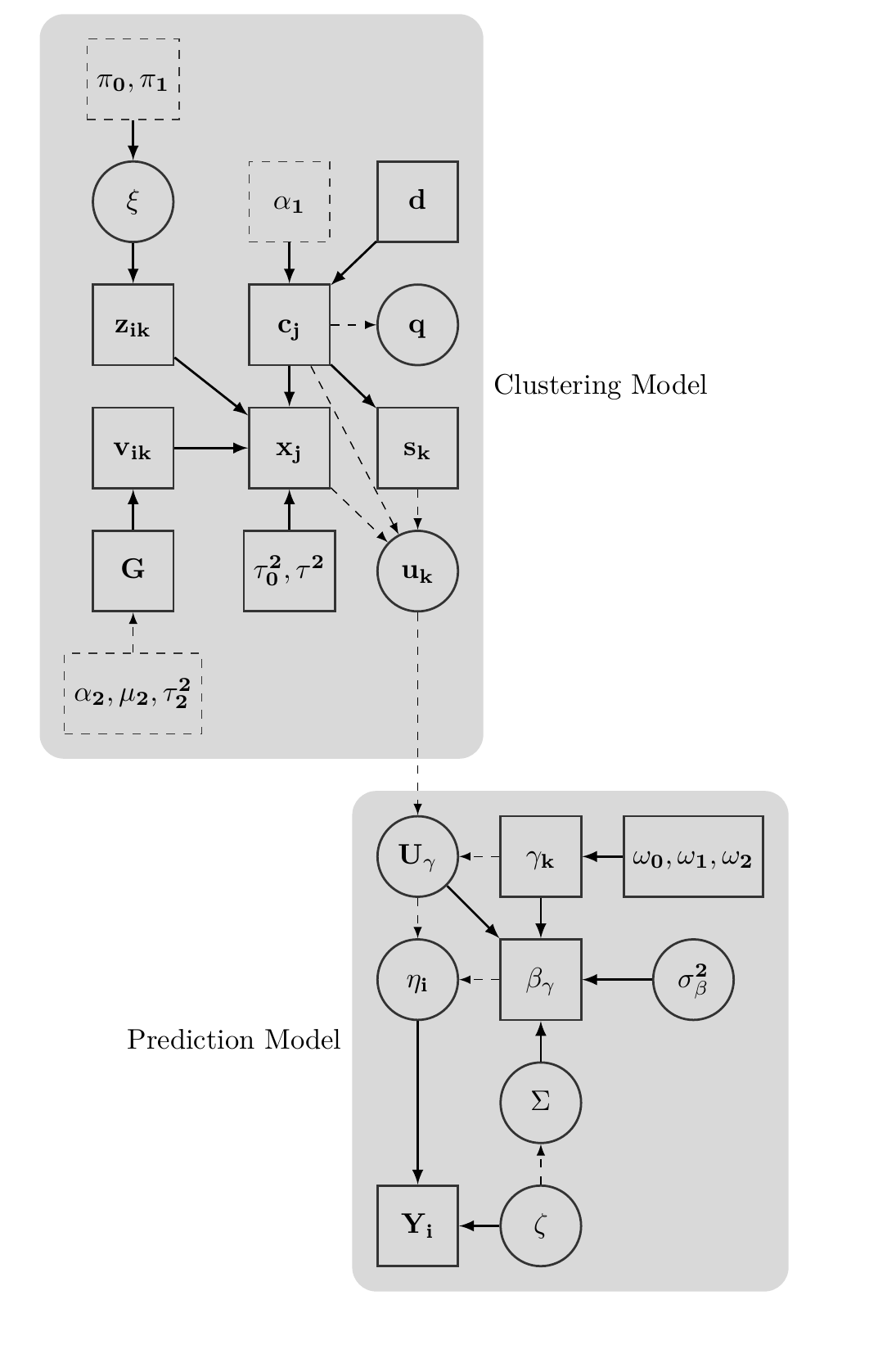}
\caption{Directed acyclic graph of the VariScan model in which the cluster representatives are chosen from the set of co-clustered covariates.  Circles represent stochastic model parameters, solid rectangles represent data and deterministic variables, and dashed rectangles represent model constants. Solid (dashed) arrows represent stochastic (deterministic) relationships.   }
\label{F:dag}
\end{center}
\end{figure}

A  schematic representation of the entire hierarchical  model involving both the clustering and prediction components is shown in Figure \ref{F:dag}.

\bigskip

\subsection{Model justification and generalizations}\label{S:justifications}

In this section, we discuss the justification, consequences, and generalizations of different aspects of the Variscan model. In particular, we investigate the appropriateness of PDPs in this application as a  tool for covariate clustering. We also discuss the choice of basis functions for the nonlinear prediction model and consider generalizations to discrete and survival outcomes.

\bigskip

\begin{figure}
\begin{center}
\includegraphics[scale=0.3]{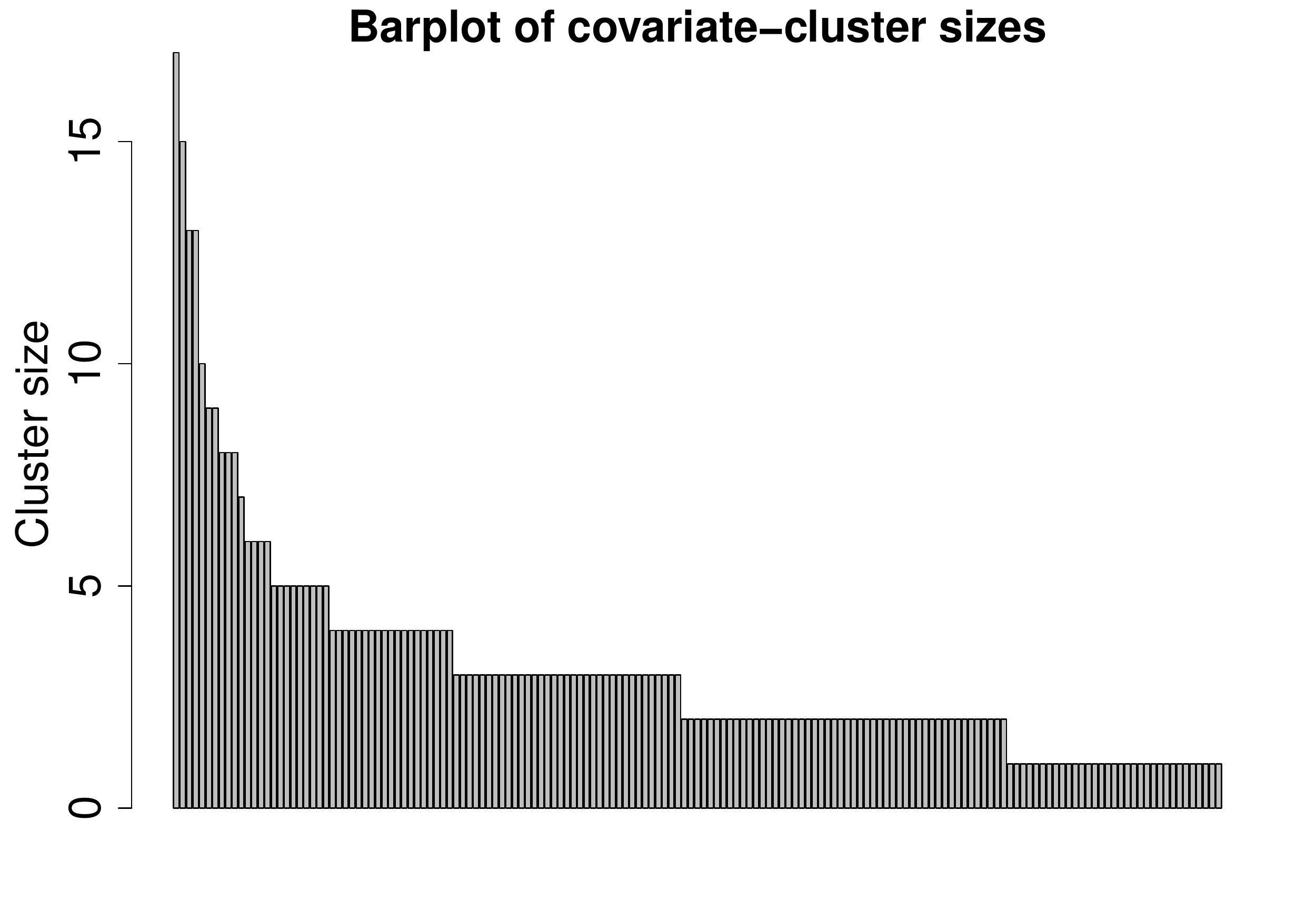}
\caption{Barchart of cluster sizes obtained by exploratory data analysis.}
\label{F:eda_barchart}
\end{center}
\end{figure}

\noindent \textbf{Empirical justification of PDPs.} We conducted an exploratory data analysis (EDA) of the gene expression levels in the DLBCL data set of \citet*{Rosenwald_etal_2002}.
Randomly selecting a set of $p=500$ probes for $n=100$ randomly chosen individuals, we iteratively applied the k-means procedure until  the covariates were grouped into fairly concordant clusters with a small overall value of $\tau^2$.
The allocation pattern depicted in Figure \ref{F:eda_barchart} is  atypical of Dirichlet processes which, as is well known among practitioners, are usually associated with relatively small number of clusters and exponentially decaying cluster sizes. Instead,
the large number of clusters  ($\hat{q}=161$) and the predominance of  small clusters suggest a non-Dirichlet model for the covariate-cluster assignments. More specifically, a PDP is favored due to the slower, power law decay in the cluster sizes typically associated with these models.

\bigskip

\noindent \textbf{Theoretical justifications for a PDP model.}
\cite{Sethuraman_1994}  derived the \emph{stick-breaking representation} for a Dirichlet process, and then \cite {Pitman_1995} extended it to  PDPs. These stick-breaking representations have the following consequences for the induced partitions. Let $\mathbb{N}$ be the set of natural numbers.
 Subject to a one-to-one transformation of the first $q$ natural numbers into $\mathbb{N}$,  the allocation variables $c_1,\ldots,c_p$ may be regarded as i.i.d.\ samples from a discrete distribution $F_{\alpha_1,d}$ on $\mathbb{N}$ with stick-breaking probabilities,  $\pi_1 = V_1$ and $\pi_h = V_h \prod_{t=1}^{h-1}(1-V_t)$ for $h = 2,3,\ldots$, where $V_h \stackrel{indep}\sim \text{beta}(1-d, \alpha_1+hd)$.
This implies that for large values of $p$   and for  clusters $k=1,\ldots,q$, the frequencies $n_k^{(p)}/p$  are approximately equal to $\pi_{h_k}$ for some distinct integers $h_1,\ldots,h_q$.

As previously mentioned, the VariScan model assumes that the base distribution $G^{(n)}$ of the PDP is the $n$-fold product measure of a univariate distribution, $G$, which follows a Dirichlet process with mass parameter $\alpha_2$. This bidirectional clustering structure has some interesting consequences.  Let $\{\phi_h\}_{h=1}^{\infty}$ be the stick-breaking probabilities associated with this  nested Dirichlet process. For two or more of the $q$ PDP clusters, the latent vectors are identical with a probability bounded above by ${q \choose 2} \cdot \bigl(\sum_{h=1}^\infty \phi_h^2\bigr)^n$. Applying the asymptotic relationship of $p$ and $q$ given in expression (\ref{q}), we find that the upper bound tends to 0 as the dataset grows, provided $p$ grows at a slower-than-exponential rate as $n$. In fact, for  $n$  as small as 50 and $p$ as small as 250,  in simulations as well as in  data analyses, we  found all the latent vectors associated with the PDP clusters to be distinct. Consequently, from a practical standpoint,  the VariScan allocations may  be regarded as clusters with  unique characteristics in even moderate-sized datasets.

Theorem \ref{Theorem:stick-breaking} below provides formal expressions for the first and second moments of the random log-probabilities of the discrete distribution $F_{\alpha_1,d}$. In conjunction with equation (\ref{q}), this result justifies the use of PDPs when the observed number of clusters is large or the cluster sizes decay slowly.  Part~\ref{T:DP_lim} provides an explanation  for the fact that   Dirichlet process allocations typically consist of a small number of clusters, only a few of which are large, with exponential decay  in the cluster sizes. Part~\ref{T:PDP_lim} suggests that in PDPs with $d>0$ (i.e., non-Dirichlet  realizations), there is a slower, power law decay of the cluster sizes as $d$ increases. Part~\ref{T:PDP_DP} indicates that for every $\alpha_1$ and $d>0$, a PDP realization $F_{\alpha_1,d}$ is thicker tailed compared to a  Dirichlet process realization, $F_{\alpha_1,0}$. See Section \ref{S_sup:stick-breaking proof} of the Appendix  for a proof.

It should be noted that the differential allocation patterns of PDPs and Dirichlet processes are well known, and has been previously emphasized by several papers, including \cite*{Lijoi_Mena_Prunster_2007a} and \cite*{Lijoi_Mena_Prunster_2007b}. However, it is difficult to come across a formal proof for this differential behavior. Although the theorem is primarily of interest when the base measure is non-atomic, it is relevant in this application because of the empirically observed uniqueness of  the latent vectors in high-dimensional~settings due to VariScan's nested structure.

\smallskip

\begin{theorem}\label{Theorem:stick-breaking}
Consider the process $\mathbb{PDP}\bigl(\alpha_1, d \bigr)$ with mass parameter $\alpha_1>0$ and discount parameter $0 \le d < 1$. Let $\psi(x)=d\log \Gamma(x)/dx$ denote the digamma function and  $\psi_1(x)=d^2\log \Gamma(x)/dx^2$ denote the trigamma function.
\begin{enumerate}

\item\label{T:PDP} For $0 < d < 1$,  the distribution $F_{\alpha_1,d} \in \mathbb{N}$ is a realization of a PDP with stick-breaking  probabilities $\pi_h$, where $h \in \mathbb{N}$. 
    However, $F_{\alpha_1,d}$ is not a Dirichlet process realization because $d \neq 0$. Then
    \begin{enumerate}
    \item\label{T:PDP_E} $E(\log \pi_h)=\psi(1-d)-\psi(\alpha_1)+\frac{1}{d}\bigl(\psi(\alpha_1/d)-\psi(\alpha_1/d+h)\bigr)$. This implies that  $\lim_{h \to \infty}E(\log \pi_h)=-\infty$.

    \item\label{T:PDP_V} $\text{Var}(\log \pi_h)=\psi_1(1-d)-\psi_1(\alpha_1)+\frac{1}{d^2}\bigl(\psi_1(\alpha_1/d)-\psi_1(\alpha_1/d+h)\bigr)$. Unlike a Dirichlet process realization, $\lim_{h \to \infty} \text{Var}(\log \pi_h)$ is finite regardless of $d>0$.

    \item\label{T:PDP_lim} For any $\alpha_1>0$ and as $h\to\infty$,
      $\log \pi_h/\log h^{-1/d} \stackrel{p}\rightarrow 1
        $ for non-Dirichlet process realizations.

    \end{enumerate}

\item\label{T:DP} For $d=0$, the distribution $F_{\alpha_1,0} \in \mathbb{N}$ is a Dirichlet process realization with stick-breaking  probabilities $\pi_h^*$ based  on $V_h^* \stackrel{iid}\sim \text{beta}(1, \alpha_1)$ for $h \in \mathbb{N}$.    Then
     \begin{enumerate}
    \item\label{T:DP_E} $E(\log \pi_h^*)=\psi(1)-\psi(\alpha_1)-h/\alpha_1$. Thus, $\lim_{h \to \infty}E(\log \pi_h^*) = -\infty$.

    \item\label{T:DP_V} $\text{Var}(\log \pi_h^*)=\psi_1(1)-\psi_1(\alpha_1)+h/\alpha_1^2$. Thus, $\lim_{h \to \infty} \text{Var}(\log \pi_h^*)=\infty$.

    \item\label{T:DP_lim} As $h\to\infty$,
     $
    \sqrt{h}\left(\frac{1}{h}\log (\pi_h^*) + 1/\alpha_1\right) \stackrel{L}\rightarrow N(0, 1/\alpha_1^2)$. This implies that as $h\to\infty$, the  random stick-breaking Dirichlet process probabilities, $\pi_h^*$, are stochastically equivalent to $e^{-h/\alpha_1}$. 
    \end{enumerate}

    \item\label{T:PDP_DP} As $h\to\infty$,
        $\sqrt{h}\left(\frac{1}{h}\log (\pi_h^*/\pi_h) + 1/\alpha_1\right) \stackrel{L}\rightarrow N(0, 1/\alpha_1^2)$.  That is,  as $h\to\infty$, the  ratios of the Dirichlet process and non-Dirichlet process stick-breaking random probabilities, $\pi_h^*/\pi_h$,  are  stochastically equivalent to $e^{-h/\alpha_1}$ for every $d>0$.  
\end{enumerate}
\end{theorem}

\smallskip

\begin{remark}
By Lemma~1 of \cite{Ishwaran_James_2003}, $\lim_{h \to \infty}E(\log \pi_h^*) = -\infty$ in Part~\ref{T:DP_E} of Theorem \ref{Theorem:stick-breaking} implies that $\sum_{h=1}^{\infty} \pi_h^*=1$  almost surely  for a Dirichlet process. A similar comment applies  in Part~\ref{T:PDP_E} for a PDP.
\end{remark}

\bigskip

\begin{figure}
\begin{center}
\includegraphics[scale=0.31]{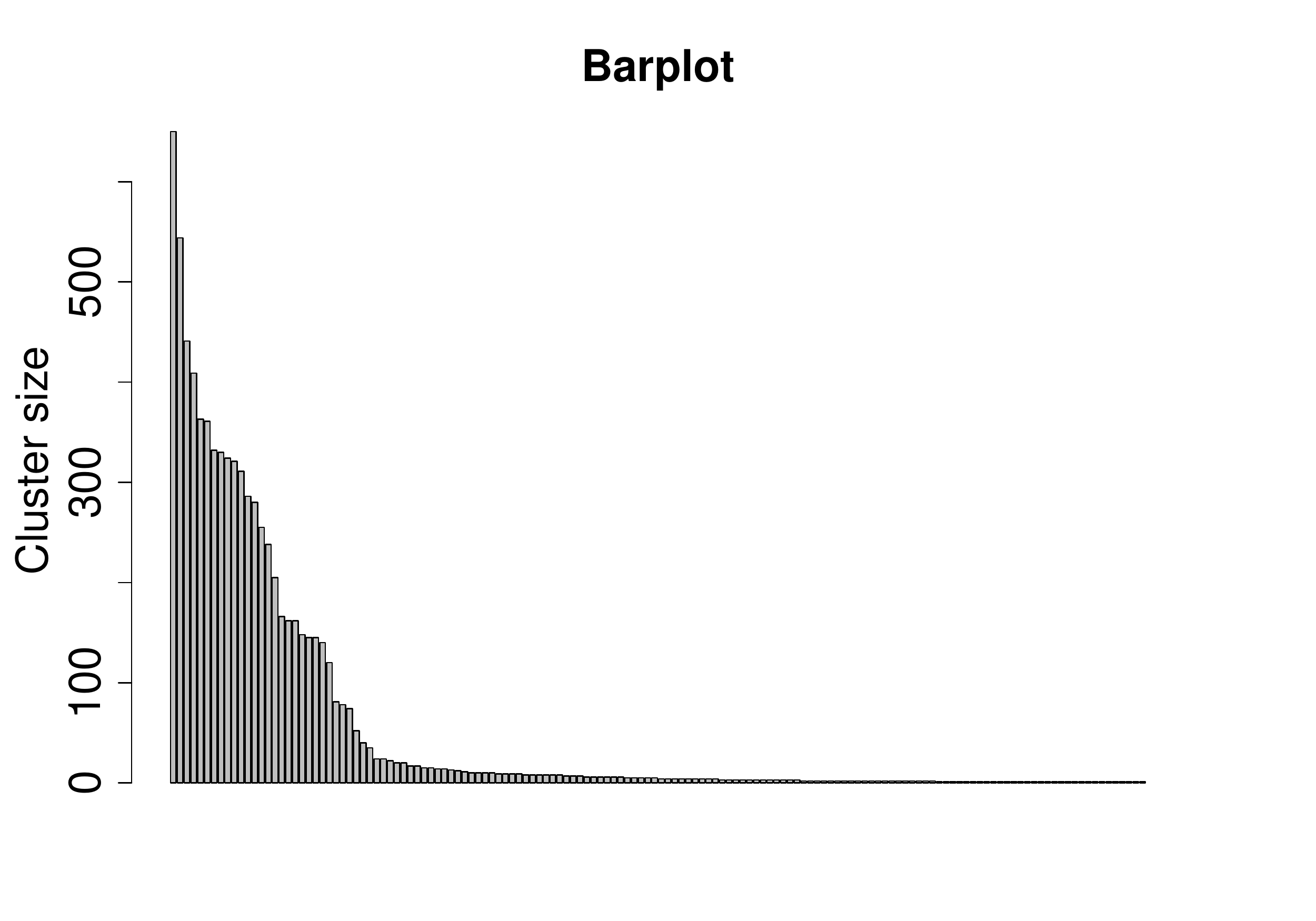}
\caption{For the DLBCL dataset, least-squares Dirichlet process configuration of the more than 14,000 latent vector elements with Bernoulli indicators equal to 1.}
\label{F:DP barchart}
\end{center}
\end{figure}

\noindent \textbf{Empirical justification of nested Dirichlet process model for the latent vector elements.}
For the DLBCL dataset, Figure \ref{F:DP barchart} presents a summary of the VariScan model estimates for the 14,000 latent vector elements with estimated Bernoulli indicators $\hat{z}_{ik}=1$. More than 87\% of the $n\hat{q}=16,500$ latent vector elements were estimated to have $\hat{z}_{ik}=1$, implying that a relatively small proportion of covariate values for the DLBCL dataset can be regarded as random noise having no clustering structure. Further details about the inference procedure are provided in Section \ref{S:inference}.  In Figure \ref{F:DP barchart}, the  small number of clusters corresponding to the large number of latent vector elements, and the sharp decline in the cluster sizes compared with Figure \ref{F:eda_barchart}, are consistent with  Dirichlet process allocation patterns. Similar results were obtained for the breast cancer data and for other genomic datasets that we have analyzed.

\bigskip

\noindent \textbf{Choice of basis functions: model parsimony versus flexibility.} \label{S:parsimony_v_flexibility}
The reliability of inference and prediction rapidly deteriorates as the number of cluster predictors and additive nonlinear components in equation (\ref{eta_i}) increases beyond a threshold value and approaches the number of subjects, $n$.  The restriction in the  prior~(\ref{gamma}) prevents  over-fitting. It ensures that the matrix $\boldsymbol{U}_{\boldsymbol{\gamma}}$, consisting of the independent regression variables, has  fewer columns than rows, and is a sufficient condition for the existence  of $({\boldsymbol{U}_{\boldsymbol{\gamma}}}'\boldsymbol{\Sigma}^{-1}\boldsymbol{U}_{\boldsymbol{\gamma}})^{-1}$ and the least-squares estimate of $\boldsymbol{\beta}_{\boldsymbol{\gamma}}$ in equation (\ref{eta_i}). In spline-based models, the relatively small number of subjects also  puts a  constraint on the order of the splines, often  necessitating  the use of linear splines with $m=1$ knot per cluster in equation (\ref{eta_i}). In the applications presented in this paper, we fixed the knot for each covariate at the sample median.

Unusually small values of $\sigma_i^2$ in equation (\ref{eta_i}) correspond to  over-fitted models, whereas unusually large values correspond to under-fitted models. Any parameters  that determine $\sigma_1^2,\ldots,\sigma_n^2$ are key, and their priors  must  be carefully chosen.
 For instance,  linear regression assumes that $\sigma_i^2=\sigma^2$. We have found that non-informative priors for $\sigma^2$ do not  work well because the optimal model sizes for  variable selection are  unknown.
 Additionally, we have found that it is  helpful to restrict the range of $\sigma^2$ based on reasonable goals for inference precision.
In the  examples discussed in this paper, we assigned the following truncated prior:
$\sigma^{-2} \sim \chi^2_{\nu} \cdot \mathcal{I}\left(0.95^{-1}/\text{Var}(\hat{\boldsymbol{y}}) < \sigma^{-2} < 0.5^{-1}/ \text{Var}(\hat{\boldsymbol{y}})\right)
$,
where the degrees of freedom $\nu$ were appropriately chosen and the  vector $\hat{\boldsymbol{y}}$ relied on EDA estimates of latent regression outcomes from a previous study or the training set individuals. The  support for $\sigma^{-2}$
approximately corresponds to the constraint, $0.5 < R^2 < 0.95$,  quantifying the effectiveness of~regression.
As Sections \ref{S:simulation} and \ref{S:benchmark_data} demonstrate, the aforementioned   strategies often result in high reliability of the response predictions.

\bigskip

\noindent \textbf{Generalizations for discrete or survival outcomes} In a general investigation, the subject-specific responses may be discrete or continuous, and/or may  be censored. 
In such cases,  the responses, denoted by $w_1,\ldots,w_n$, can be modeled as deterministic transformations of random variables~$R_i$ from an exponential family distribution. The Laplace approximation  \citep{Harville_1977} transforms each  $R_i$ into a Gaussian \textit{regression outcome},~$y_i$, that can be modeled using our Variscan model proposed above.
The details of the calculation are as follows.
For a set of functions~$f_i$, we  assume that $w_i = f_i(R_i)$ and  density function
$[R_i \mid \varrho_{i}, \varsigma ] = r(R_i,\varsigma)\cdot \exp\left(
\frac{R_i\,\varrho_{i}-b(\varrho_{i})}{a(\varsigma)}\right)$,
where $r(\cdot)$ is a non-negative function, $\varsigma$ is a dispersion parameter,   $\varrho_{i}$ is the canonical parameter, and $[\cdot]$ represents  densities with respect to a dominating measure.
The Laplace approximation  relates the $R_i$'s to Gaussian regression outcomes:
$y_{i} = \eta_{i} + \frac{\partial \eta_{i}}{\partial \mu_{i}}\cdot(R_{i} - \mu_{i})$ is approximately  $N\left(\eta_{i},\sigma_i^2\right)$
with  precision $\sigma_i^{-2}=\{b^{''}(\mu_{i})\}^{-1}\left(\partial
\mu_{i}/\partial \eta_{i}\right)^2$.
For an appropriate link function $g(\cdot)$, the mean $\eta_{i}$ equals $g(\mu_{i})$.
Gaussian, Poisson, and binary responses are applicable in this setting.
Accelerated failure time (AFT) censored outcomes \citep{Buckley_James_1979, Cox_Oakes_1984} also fall into this modeling framework.

The idea of using a Laplace-type approximation for exponential families is not new. Some early examples in  Bayesian settings include \cite{Zeger_Karim_1991}, \cite{Chib_Greenberg_1994},  and \cite{Chib_Greenberg_Winkelmann_1998}.
For linear regression, the  approximation is exact
with $y_{i} = R_{i}$.
The Laplace approximation
is not restrictive even when it is approximate; for example, MCMC proposals for the model parameters can be filtered through a Metropolis-Hastings step to obtain  samples from the target posterior. Alternatively, inference strategies relying on normal mixture representations  through  auxiliary variables could be used to relate the $R_i$'s to the $y_i$'s. For instance, \cite{Albert_Chib_1993} used truncated normal sampling to obtain a probit model for binary responses, and \cite{HolmesHeld2006}
 utilized a scale mixture representation of the normal distribution \citep{Andrews_Mallows_1974,West_1987} to implement logistic regression using latent variables.

\bigskip

 \section{Posterior inference}\label{S:inference}

 Starting with an initial configuration obtained by a na\"{i}ve, preliminary analysis, the model parameters are iteratively updated by MCMC methods.
 Due to the intensive nature of the posterior inference, the analysis is performed in two stages, with cluster detection followed by predictor discovery:

 \smallskip

 \begin{enumerate}
 \item[\textbf{Stage 1}] Focusing  only on the covariates and ignoring the responses:

 \smallskip

      \begin{enumerate}
    \item[\textit{Stage 1a}] The allocation variables,  latent vector elements, and binary indicators are iteratively updated until the MCMC chain converges. Monte Carlo estimates are computed for the posterior probability of clustering for each pair of covariates. Applying the technique of  \cite{Dahl_2006}, these pairwise probabilities are used to compute a point estimate, called the \textit{least-squares  allocation}, for the allocation~variables.
        Further details of the MCMC procedure are provided in Sections \ref{S:MCMC_c} and \ref{S:MCMC_v} of the Appendix.

 \smallskip

    \item[\textit{Stage 1b}] Conditional on the least-squares  allocation being the true clustering of the covariates, a second MCMC sample of the latent vector elements and binary indicators is generated. Again applying  the technique of \cite{Dahl_2006}, we compute  a point estimate, called the \textit{least-squares configuration}, for the latent vector elements and  binary  indicators.
  \end{enumerate}

 \bigskip

 \item[\textbf{Stage 2}]  Conditional on the least-squares allocation and  least-squares configuration, and focussing on the responses,  the cluster predictors and  latent regression outcomes, if any, are generated to obtain a third MCMC sample. The MCMC sample is post-processed to predict the responses for out-of-the-bag test set individuals. The interested reader is referred to Sections \ref{S:MCMC_gamma}, \ref{S:MCMC_y} and \ref{S:test_case_prediction} of the Appendix for  details.
 \end{enumerate}

As a further benefit of  a coherent model for the covariates, VariScan is able to perform model-based imputations of any missing subject-specific covariates as part of the MCMC procedure.

\bigskip

\section{Clustering consistency}\label{S:clusters}

As  mentioned in Section \ref{S:covariates}, the latent vectors associated with  two or more PDP clusters may be identical under the VariScan model, but this probability becomes vanishingly small as $n$ grows. Consequently, for practical purposes, the VariScan   allocations may be interpreted as distinct, identifiable clusters in even moderately large datasets. In order to study the reliability of VariScan's clustering procedure in our targeted Big Data applications, we
 make large-sample theoretical comparisons between the VariScan model's cluster allocations and the true allocations of a hypothetical covariate generating process.

 In the general problem of using mixture models to allocate $p$ objects to an unknown number of clusters, the problem of non-identifiability and redundancy of the detected clusters has been extensively documented in Bayesian and frequentist applications \citep[e.g., see][]{Fruhwirth-Schnatter_2006}. Some partial solutions are available in the Bayesian literature. For example, in finite mixture models, rather than assuming exchangeability of the mixture component parameters, \cite{Petralia_etal_2012} regard them as draws
from a repulsive process, leading to fewer, better separated and more interpretable
clusters. \cite{Rousseau_Mengersen_2011} show that  a carefully chosen prior leads to asymptotic emptying of the
redundant components in  over-fitted finite mixture models.
The underlying strategy of these procedures is that they focus on detecting the correct number of clusters rather than the correct allocation of the $p$~objects.

In contrast to the non-identifiability of the detected clusters in fixed $n$ settings, Theorem \ref{Thm:consistency} establishes the interesting fact that, when $p$ and $n$ are both large, a fixed set of  covariates that (do not) co-cluster under the true process,  also (do not) asymptotically co-cluster under the  posterior. The key intuition is that, as with most mixture model applications, when $n$-dimensional objects are clustered and $n$ is small, it is possible for the clusters to be erroneously placed too close together even if $p$ is large. On the other hand, if $n$ is  allowed to grow with $p$, then objects in $\mathcal{R}^n$  eventually become well separated. 
Consequently,  for $n$ and $p$ large enough, the VariScan method is able to infer the true clustering for a fixed subset of the $p$ covariate columns. In the sequel, using synthetic datasets  in Section~\ref{S:simulation2}, we exhibit the high accuracy of VariScan's clustering-related inferences.

  \bigskip

  \paragraph{\textbf{The true model.}} The VariScan model's   exchangeability assumption for the $p$ covariates stems from our belief in the existence of a true, unknown de Finetti density in $\mathcal{R}^n$ from which the  column vectors arise as a random sample. In particular, for any given $n$ and $p$, we make the following assumptions about the true covariate-generating process:

\begin{enumerate}[(a)]
\item\label{true:first} The column vectors $\boldsymbol{x}_1,\ldots,\boldsymbol{x}_p$ are a random sample of size $p$ from an $n$-variate distribution $P_0^{(n)}$ convolved with  $n$-variate, independent-component Gaussian errors.
      \item The true distribution $P_0^{(n)}$ is discrete in the space $\mathcal{R}^n$. Let the $n$-dimensional  atoms of $P_0^{(n)}$ be denoted by $\boldsymbol{v}^{(0)}_t=(v^{(0)}_{1t},\ldots,v^{(0)}_{nt})'$ for positive integers $t$.
          \item\label{true.alloc} Due to the discreteness of distribution $P_0^{(n)}$, there exist true allocation variables, $c_1^{(0)},\ldots,c_p^{(0)}$, mapping the $p$ covariates to   distinct atoms of  $P_0^{(n)}$. For subjects $i=1,\ldots,n$, and columns $j=1,\ldots,p$, the covariates are then distributed as
              \begin{equation}
              x_{ij} \mid c_j^{(0)} \stackrel{indep}\sim N(v^{(0)}_{i\, c_j^{(0)}}, \tau_0^2), \label{true.lik.x}
              \end{equation}
\item The $n$-variate atoms of  distribution $P_0^{(n)}$ are  i.i.d.\ realizations of the $n$-fold product measure of a univariate distribution, $G_0$. Consequently, the atom elements are $v^{(0)}_{it}\stackrel{i.i.d.}\sim G_0$ for  $i=1,\ldots,n$, and  $j=1,\ldots,p$.

    \item\label{true:last} The true distribution $G_0$ is non-atomic and has compact support on the real line.
\end{enumerate}

\bigskip

Let $\mathcal{L}=\{j_1,\ldots, j_L\} \subset \{1,\ldots,p\}$ be a fixed subset of $L$ covariate indexes. Given a vector of inferred allocations $\boldsymbol{c}=(c_1,\ldots,c_p)$, we quantify the inference accuracy  by the \textit{proportion of correctly clustered covariate pairs}:
        \begin{equation}
        \varkappa_{\mathcal{L}}(\boldsymbol{c}) = \frac{1}{{L \choose 2}} \sum_{j_1 \neq j_2 \in \mathcal{L}} \mathcal{I}\biggl(\mathcal{I}(c_{j_1}=c_{j_2})=\mathcal{I}(c_{j_1}^{(0)}=c_{j_2}^{(0)})\biggr). \label{varkappa}
        \end{equation}
        A  value near 1 indicates the high  accuracy of inferred allocations $\boldsymbol{c}$ for the set $\mathcal{L}$. Notice that the measure $\varkappa_{\mathcal{L}}(\boldsymbol{c})$ is invariant to  permutations of the clusters labels. This is desirable because the  labels  are arbitrary.

\bigskip

\begin{theorem}\label{Thm:consistency}
Denote the covariate matrix by $\boldsymbol{X}_{np}$. In addition to assumptions (\ref{true:first})--(\ref{true:last}) about the true covariate-generating process, suppose that the true standard deviation  $\tau_0$ in equation (\ref{true.lik.x}) is bounded below by $\tau_*$, the small, positive   constant postulated in Section~\ref{S:covariates} as a lower bound for the Variscan model parameters, $\tau_1$ and $\tau$.

Let $\mathcal{L}=\{j_1,\ldots, j_L\} \subset \{1,\ldots,p\}$ be a fixed subset of $L$ covariate indexes.  Then there exists an increasing sequence of numbers $\{p_n\}$ such that, as $n$ grows and provided  $p>p_n$, the VariScan clustering inferences for the covariate subset $\mathcal{L}$ are aposteriori consistent. That is,
\[
\lim_{\substack{n \to \infty \\ p> p_n}} P\bigl[\varkappa_{\mathcal{L}}(\boldsymbol{c})=1 \mid \boldsymbol{X}_{np}\bigr] \to 1.
\]

\end{theorem}

\bigskip

See Section \ref{S:proof of Thm:consistency} of the Appendix for a proof.
The   result relies on  non-trivial extensions, in several directions, of the  important theoretical insights provided by \citep{Ghosal_Ghosh_Ramamoorthi_1999}.
Specifically, it extends Theorem 3 of \cite{Ghosal_Ghosh_Ramamoorthi_1999} to densities on $\mathcal{R}^n$ arising as convolutions of vector locations with  errors distributed as  zero-mean finite normal mixtures. 
\bigskip

\section{Simulation studies}

\smallskip

\subsection{Cluster-related inferences}\label{S:simulation2}

\begin{figure}
\centering
\includegraphics[scale=0.4]{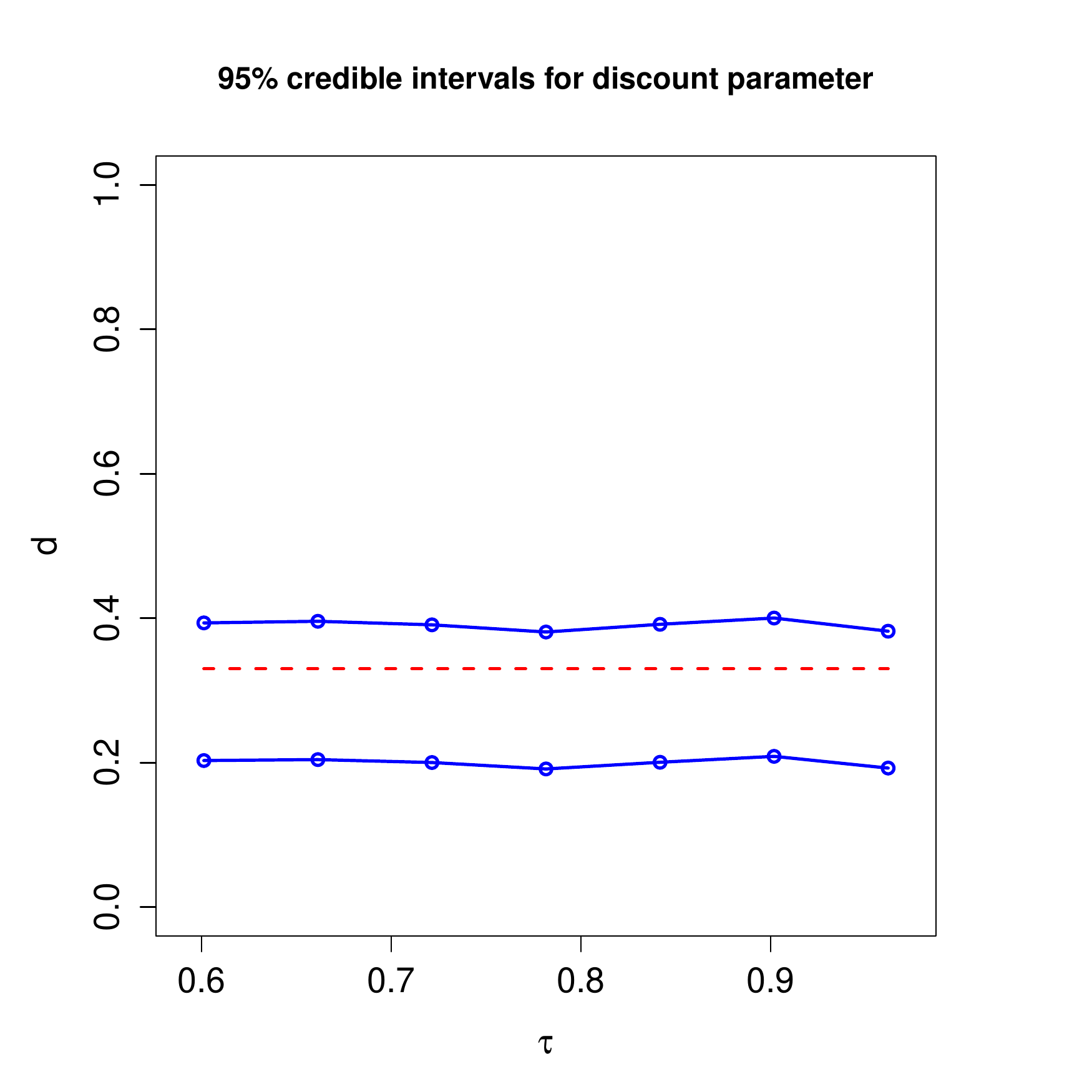}
\caption{ 95\% posterior credible intervals for the discount parameter, $d$ for different values of $\tau_0$. The true value, $d_0$, is shown by the red dashed line.}
\label{F:d}
\end{figure}

 We investigate VariScan's accuracy as a clustering procedure using artificial datasets for which the true clustering pattern is known. We  simulated the covariates for $n=50$ subjects  and $p=250$ genes from a discrete distribution convolved with Gaussian noise, and compared the  co-clustering posterior probabilities of the $p$ covariates with the  truth. The parameters of the true model were chosen to approximate match the corresponding estimates for the DLBCL dataset of \cite{Rosenwald_etal_2002}. Specifically, for each of 25 synthetic datasets, and for the true model's parameter $\tau_0$ in Theorem~\ref{Thm:consistency} belonging to the range $[0.60, 0.96]$, we generated the following quantities to obtain the matrix $\boldsymbol{X}$ in Step \ref{X_step} below:

\begin{enumerate}

 \item \textbf{True allocation variables:} We generated  $c_1^{(0)},\ldots,c_p^{(0)}$ as the partitions induced by a PDP with true discount parameter $d^{(0)}=0.33$ and mass parameter $\alpha_1=20$. The true number of clusters, $Q_0$, was thereby computed for this non-Dirichlet allocation.

\item \textbf{Latent vector elements:} For $i=1,\ldots,n$ and $k=1,\ldots,Q_0$, elements $v_{ik}^{(0)} \stackrel{iid}\sim G_0$, where $G_0 \sim \mathcal{DP}(\alpha_2)$,
with mass $\alpha_2=10$ and uniform base distribution $U_0$  on the interval $[1.4,2.6]$.

\item\label{X_step} \textbf{Covariates:} $x_{ij} \stackrel{indep}\sim N(v_{ic_j}^{(0)}, \tau_0^2)$ for $i=1,\ldots,n$ and $j=1,\ldots,p$.

\end{enumerate}

  No responses were generated in this study. 
   Applying  the general technique of \cite{Dahl_2006} developed for Dirichlet process models, we computed  a point estimate for the  allocations, called the \textit{least-squares configuration}, and denoted by $\hat{c}_1,\ldots,\hat{c}_p$. For the full set of covariates, we  estimated the accuracy of the least-squares allocation by the \textit{estimated proportion of correctly clustered covariate pairs},
        \[
        \hat{\varkappa} = \frac{1}{{p \choose 2}}  \sum_{j_1 \neq j_2 \in \{1,\ldots,p\}} \mathcal{I}\biggl(\mathcal{I}(\hat{c}_{j_1}=\hat{c}_{j_2})=\mathcal{I}(c_{j_1}^{(0)}=c_{j_2}^{(0)})\biggr).
        \]
        A high value of $\hat{\varkappa}$ is indicative of VariScan's high clustering accuracy for all $p$ covariates.

       For each value of $\tau_0$, the second column of Table \ref{T:varkappa} displays the percentage $\hat{\varkappa}$ averaged over the 25 independent replications. We find that, for each $\tau_0$,  significantly less than 5  pairs  were incorrectly clustered out of the ${250 \choose 2}=$ 31,125 different covariate pairs, and so $\hat{\varkappa}$ was significantly greater than 0.999. The posterior inferences appear to be robust to large noise levels, i.e., large values of $\tau_0$. For every  dataset, $\hat{q}$, the estimated number of clusters  in the least-squares  allocation was exactly equal to $Q_0$, the true number of~clusters. Recall that the non-atomicity of true distribution $G_0$ is a  sufficient condition of  Theorem~\ref{Thm:consistency}. Although the condition is not satisfied in this setting, we nevertheless obtained highly accurate clustering-related inferences for the full set of $p=250$~covariates.

       Accurate inferences were also obtained for the PDP discount parameter, $d \in [0,1)$. Figure \ref{F:d} plots the 95\% posterior credible intervals for $d$ against different values of $\tau_0$. The posterior inferences are substantially more precise than the prior and each interval contained the true value,~$d_0=0.33$. Furthermore, in spite of being assigned a prior probability of 0.5, there is no posterior mass allocated to Dirichlet process models.
The ability of VariScan to discriminate between PDP and Dirichlet process models was evaluated using the log-Bayes factor, $
    \log\left(P[d>0|\boldsymbol{X}]/P[d=0|\boldsymbol{X}]\right)$. With $\Theta^*$ representing all the parameters except $d$, and applying Jensen's inequality, the log-Bayes factor exceeds $E\left(\log\left(\frac{P[d>0|\boldsymbol{X},\Theta^*]}{p[d=0|\boldsymbol{X},\Theta^*]} \right) \mid \boldsymbol{X} \right)$, which (unlike the log-Bayes factor) can be estimated using just the post--burn-in MCMC sample. For each $\tau_0$, the third column of  Table \ref{T:varkappa} displays 95\% posterior credible intervals for this
     lower bound.
 The Bayes factors are significantly greater than $e^{10}=22,026.5$ and are overwhelmingly in favor of PDP~allocations, i.e.,  the true model.

 {\small

\begin{table}
\begin{center}
\renewcommand{\arraystretch}{1}
\begin{tabular}{ c   | c |c  }
\hline\hline
 \textbf{True $\tau_0$} &\textbf{Percent $\hat{\varkappa}$} &\textbf{95\% C.I.\ for lower  }\\
 & &\textbf{bound of  log-BF} \\
\hline
0.60 &99.984 (0.000) &(11.05, 11.10)\\
0.66 &99.978 (0.000) &(11.17, 11.25)\\
0.72 &99.976 (0.000) &(10.89, 10.98)\\
0.78 &99.973 (0.001) &(10.23, 10.31)\\
0.84 &99.971 (0.000) &(10.86, 10.93)\\
0.90 &99.960 (0.000) &(11.88, 11.94)\\
0.96 &99.941 (0.001) &(10.49, 10.56)\\
\hline\hline
\end{tabular}
\end{center}
\caption{For different values of simulation parameter $\tau_0$, column 2 displays the proportion of correctly clustered covariate pairs, with the standard errors for the 25 independent replications shown in  parentheses. Column 3 presents 95\% posterior credible intervals for the lower bound of the log-Bayes factor of PDP models relative to Dirichlet process models. See the text for further explanation.}\label{T:varkappa}
\end{table}
}

\subsection{Prediction accuracy}\label{S:simulation}

We evaluate the operating characteristics of our methods using a simulation study based
on the  DLBCL dataset of \cite{Rosenwald_etal_2002}. To generate the simulated data, we selected $p=500$ genes from the original gene expression dataset of 7,399 probes, as detailed below:

\begin{enumerate}

\item Select $10$ covariates with pairwise correlations less than 0.5 as the true predictor set, $\mathcal{S} \subset\{1,\ldots,500\}$, so that $|\mathcal{S}|=10$.

\item For each value of $\beta^*\in \{0.2, 0.6, 1.0\}$:

\begin{enumerate}

    \item For subjects $i=1,\ldots,100$, generate failure times $t_i$ from distribution $ \mathcal{E}_i$, denoting the exponential distribution with mean~$\exp(\beta^*  \sum_{j\in \mathcal{S}}  x_{ij})$. Note that the model used to generate the outcomes differs from  VariScan assumption~(\ref{eta_i}) for the log-failure times.

\item For 20\% of individuals, generate their censoring times as follows: $u_i \sim$ $\mathcal{E}_i \cdot \mathcal{I}(u_i < t_i)$. Set the survival times of these individuals to $w_i=\log u_i$ and their failure statuses to $\delta_i=0$.

\item For the remaining individuals, set $w_i = \log t_i$ and $\delta_i=1$.

\end{enumerate}

\item Randomly assign 67 individuals to the training set and the remaining 33 individuals to the test set.

\item  Assuming the AFT survival model, apply the  VariScan procedure with linear splines and $m=1$ knot per spline. Choose a single covariate from each cluster as the representative as described in Section \ref{S:predictors}. Make posterior inferences using the training data and predict the outcomes for the test~cases.

\end{enumerate}

We analyzed the
 same set of simulated data using six other techniques  for gene selection with survival outcomes: lasso \citep{Tibshirani_1997}, adaptive lasso \citep{Zou_2006}, elastic net \citep{Zou_Trevor_2005}, $L_2$-boosting \citep{Hothorn_Buhlmann_2006}, random survival forests \citep{Ishwaran_etal_2010}, and supervised principal components \citep{Bair_Tibshirani_2004}, which have been implemented in the R packages glmnet, mboost, randomSurvivalForest, and superpc. The ``RSF-VH'' version of the  random survival forests procedure was chosen because of its success in high-dimensional~problems. The selected techniques are  excellent examples  of the three categories of approaches  for small $n$, large $p$ problems (variable
selection, nonlinear prediction, and regression
based on lower-dimensional projections) discussed  in Section~\ref{S:introduction}.
We repeated this procedure over fifteen independent replications.

We compared the  prediction errors of the methods using the \textit{concordance error rate}, which is defined as $1-C$, where $C$ denotes the c index  of  \cite{Harrell_etal_1982}. Let the set of ``usable'' pairs of subjects be $\mathcal{U} = \{(i,j): w_i < w_j, \delta_i=1\} \cup \{(i,j): w_i = w_j, \delta_i\neq \delta_j\}$. The concordance error rate of a procedure is  \citep{May_etal_2004}:
 $
 1 -  C  = \frac{1}{|\mathcal{U}|}\sum_{(i,j) \in \mathcal{U}} \mathcal{I}(\tilde{w}_i \ge \tilde{w}_j) - \frac{1}{2|\mathcal{U}|}\sum_{(i,j) \in \mathcal{U}} \mathcal{I}(\tilde{w}_i = \tilde{w}_j)
$,
where $\tilde{w}_i$ is the predicted response of subject $i$. For example, for the VariScan procedure applied to analyze AFT survival outcomes, the predicted responses are $\tilde{w}_i=\exp(\tilde{y}_i)$, where
 $\tilde{y}_i$ are the predicted regression outcomes. 

 \begin{figure}
\begin{center}
\includegraphics[scale=0.31]{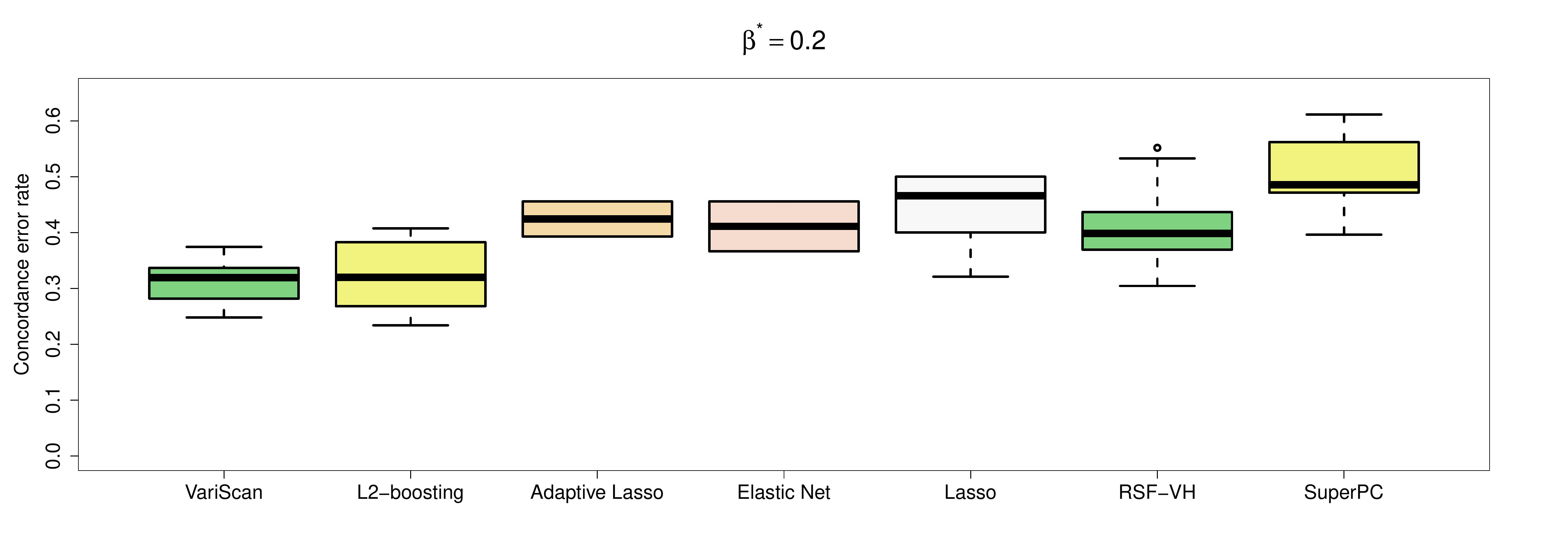}
\includegraphics[scale=0.31]{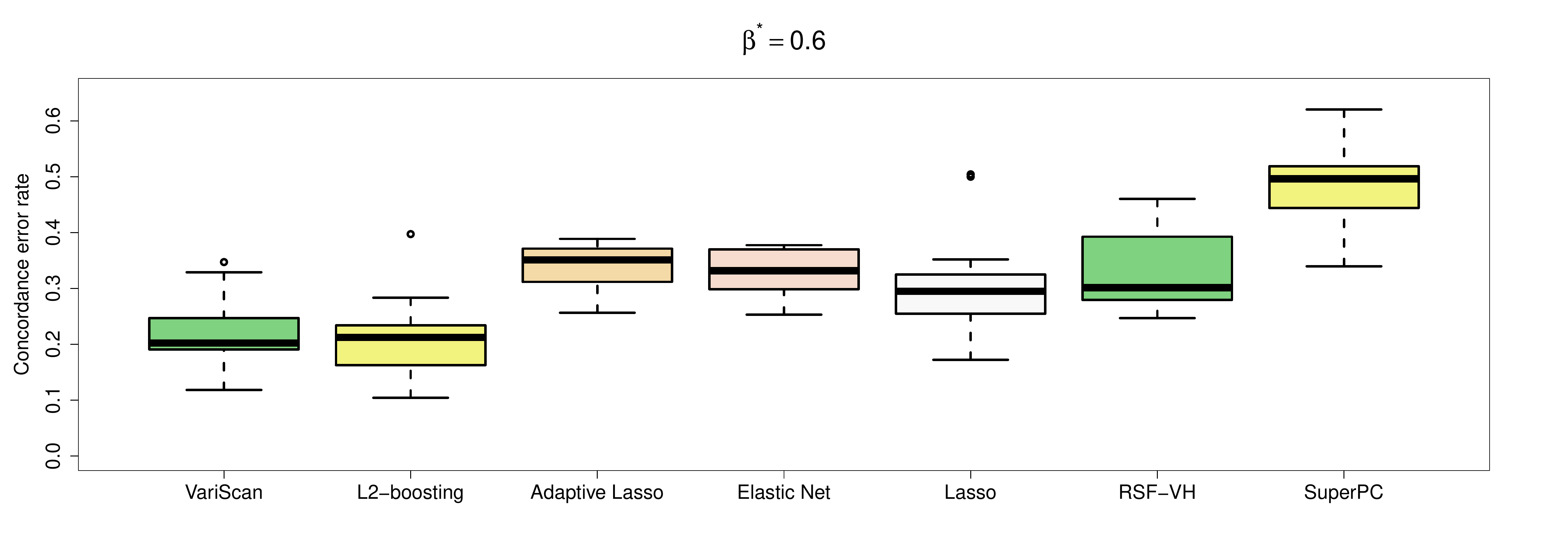}
\includegraphics[scale=0.31]{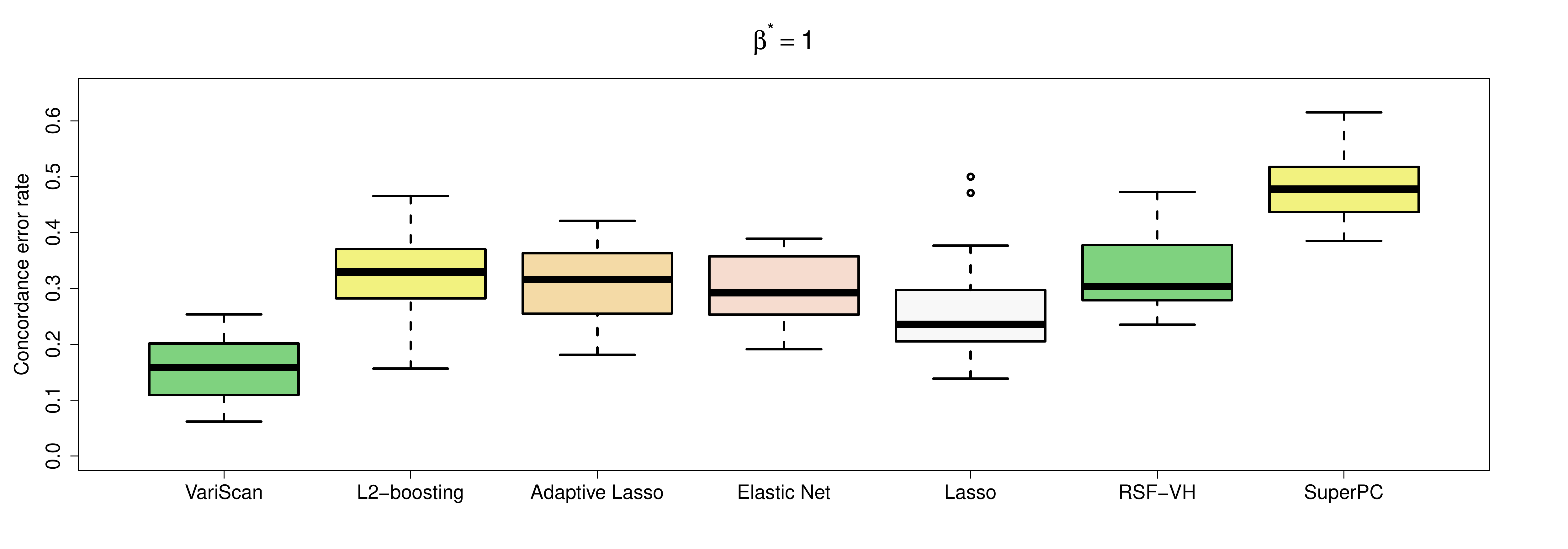}
\caption{Side-by-side boxplots comparing the  percentage concordance error rates of the different techniques in the simulation study.}
\label{F:C2_simulation}
\end{center}
\end{figure}

The concordance error rate measures a procedure's probability of incorrectly ranking the failure times of two randomly chosen individuals. The accuracy of a procedure is inversely related to its concordance error rate. The measure is especially useful for comparisons because it does not rely on the survivor function, which is estimable by VariScan, but not by some of the other procedures.
 Figure~\ref{F:C2_simulation} depicts  boxplots of the concordance error rates of the  procedures sorted by increasing order of prediction accuracy. 
 We find that as $\beta^*$ increases, the concordance error rates progressively decrease for most procedures, including VariScan.  For larger $\beta^*$, the error rates for VariScan are significantly lower than the error rates for the other~methods.

\begin{figure}
\begin{center}
\includegraphics[scale=0.5]{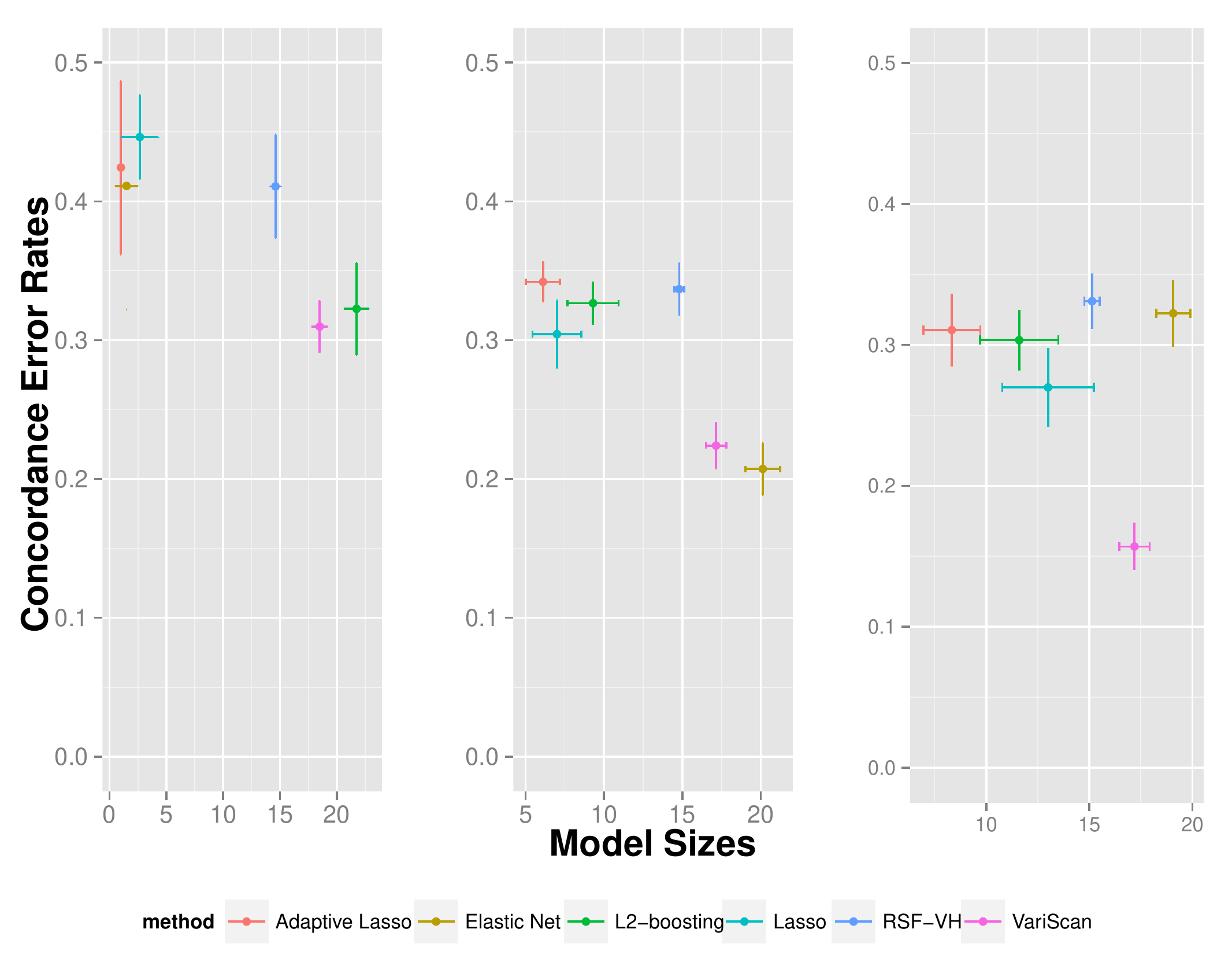}
\caption{Plot of concordance error rates versus model sizes for the competing methods along with the standard errors (shown by whiskers).
The left, middle and right respectively correspond to effect size $\beta^*$ equal to 0.2, 0.6, and 1.}
\label{F:sims}
\end{center}
\end{figure}

In order to facilitate a more systematic evaluation, we have plotted in Figure~\ref{F:sims} the error rates versus model sizes for the different methods, thereby providing a joint examination of model parsimony and prediction. To aid a visual interpretation, we did not include the  supervised principal components method, since it performs the worst in terms of prediction and detects models that are two to four fold larger than $L_2$-boosting, which typically produces the largest models among the depicted methods. The three panels correspond to increasing effect size, $\beta^*$.  A few facts are evident from the plots. VariScan seems to balance sparsity and prediction the best for all values of $\beta^*$, with its performance increasing appreciably with $\beta^*$. Penalization approaches such as lasso, adaptive lasso, and elastic net produce sparser models but have lower prediction accuracies. $L_2$-boosting is comparable to Variscan in terms of prediction accuracy, but detects larger models for the lower effect sizes (left and middle panel); Variscan is the clear winner for the largest effect size (right panel). Additionally, especially for the largest $\beta^*$, we observe substantial variability between the simulation runs for the penalization approaches,  as reflected by the large standard errors. Further simulation study comparisons of VariScan and the  competing approaches are presented in Section \ref{SA:prediction_simulation} of the Appendix.

\textbf{Nonlinearity measure.} Unlike some existing approaches,  VariScan is able to measure the degree of nonlinearity in the  relationships between the responses and covariates. For example, we could define \textit{nonlinearity measure} $\mathcal{N}$ as the posterior expectation,
\begin{equation}
\mathcal{N} = E\bigl( \frac{\omega_2}{\omega_1+\omega_2} | \boldsymbol{w},\boldsymbol{X}\bigr). \label{N}
\end{equation}
This represents the posterior odds that a hypothetical, new  cluster is a non-linear predictor in equation~(\ref{eta_i}) rather than a simple linear regressor.  
A value of $\mathcal{N}$ close to 1 corresponds to predominantly nonlinear associations between the responses and their predictors.

Averaging over the 15 independent replications of the simulation,  as $\beta^*$ varied over the set $\{0.2, 0.6, 1.0\}$, the estimates of the nonlinearity measure $\mathcal{N}$ defined in equation~(\ref{N}), were  0.72, 0.41, and 0.25, respectively. The corresponding standard errors were 0.04, 0.07, and 0.06. This indicates that on the scale of the simulated log--failure times, simple linear regressors are increasingly preferred to linear splines  as the signal-to-noise ratio, quantified by $\beta^*$, increases. Such interpretable measures of nonlinearity are not provided by the competing methods.

\bigskip

\begin{center}
\section{Analysis of benchmark data sets} \label{S:benchmark_data}
\end{center}

Returning
 to the two publicly available datasets of Section \ref{S:introduction}, we chose  $p=500$ probes for further analysis. For  the DLBCL dataset of  \citet*{Rosenwald_etal_2002}, we randomly selected 100 out of the 235 individuals who had  non-zero survival times. Of the individuals selected, 50\% had censored failure times. For the breast cancer dataset of \citet*{vantVeer_2002}, we analyzed the 76 individuals with non-zero survival times, of which 44 individuals (57.9\%)  had censored failure times.

We performed 50  independent replications of the three steps that  follow. \textit{(i)} We randomly split the data into training and test sets in a 2:1 ratio.  \textit{(ii)} We analyzed the survival times and $p=500$ gene expression levels of the training cases using the techniques VariScan, lasso, adaptive lasso, elastic net, $L_2$-boosting, random survival forests, and supervised principal components. \textit{(iii)}~The different techniques were used to predict the test~case outcomes. For the VariScan procedure, a single covariate from each cluster was chosen to be the cluster representative.

The number of clusters  for the least-squares  allocation of covariates, $\hat{q}$, computed in Stage 1a of the analysis, were  165 and 117 respectively for the DLBCL and the breast cancer  datasets. The nonlinearity measure $\mathcal{N}$ estimates were  0.97 and 0.75 respectively with small standard errors. This   indicates that the responses in both  datasets,  but  especially    in the DLBCL dataset,  have predominantly nonlinear relationships with the predictors. In spite of being assigned a prior probability of 0.5, the estimated posterior probability of the Dirichlet process model (corresponding to  discount parameter $d=0$) was exactly 0 for both  datasets, justifying the PDP-based allocation scheme.

\begin{figure}
\begin{center}
\includegraphics[scale=0.31]{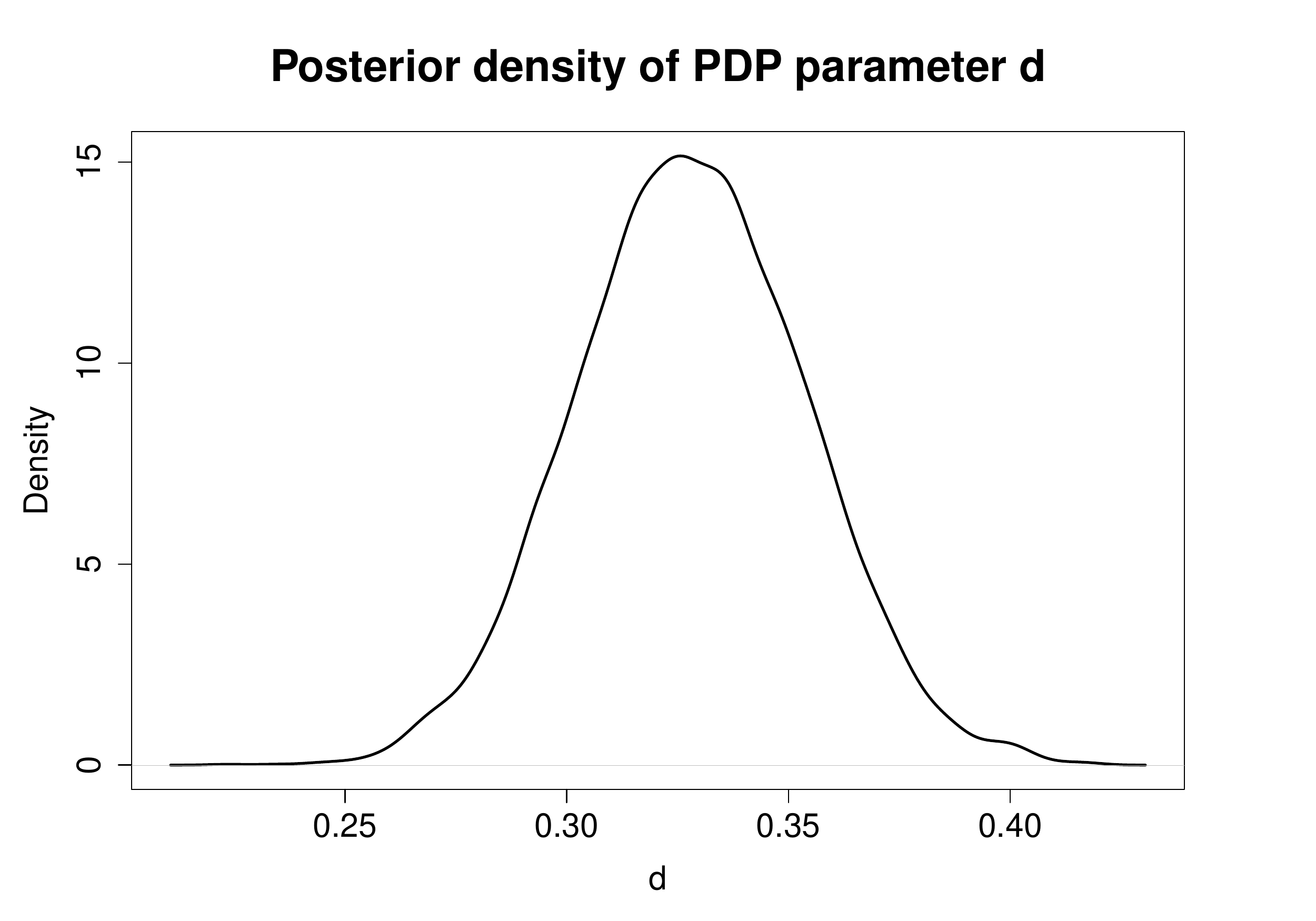}
\includegraphics[scale=0.31]{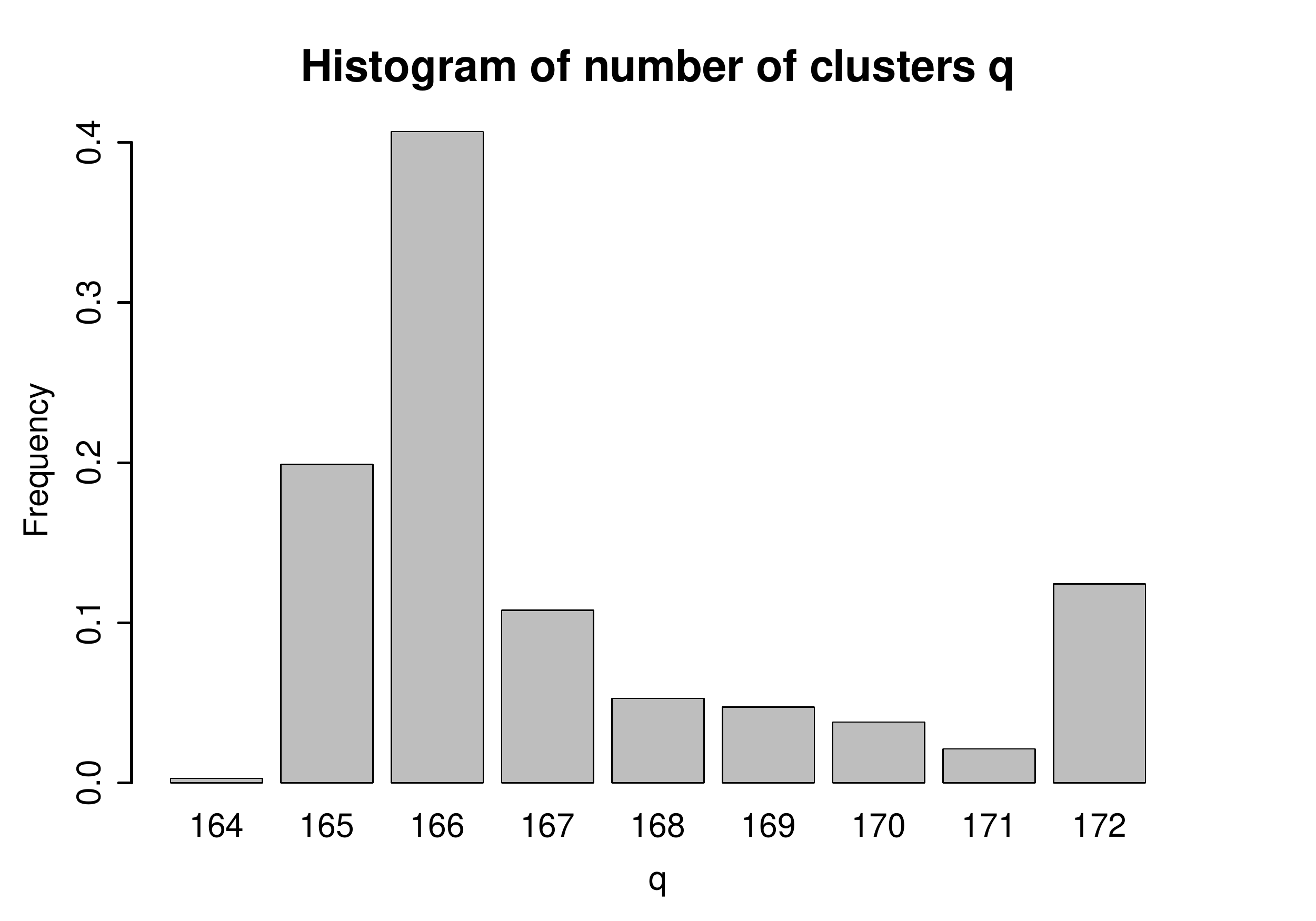}
\includegraphics[scale=0.31]{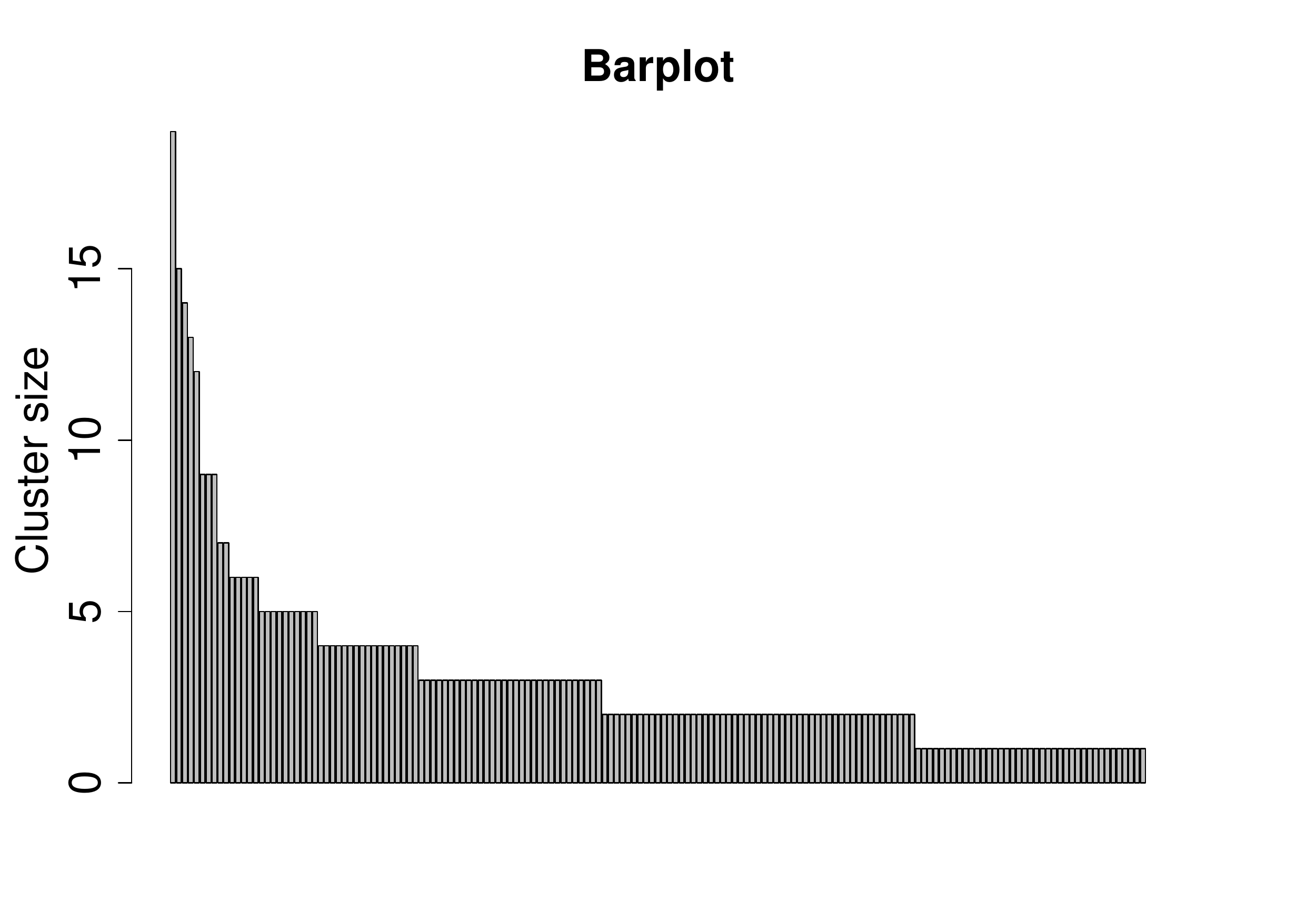}
\caption{Posterior summaries for the DLBCL dataset. The top panels and the lower panel summarize the least-squares covariate-to-cluster PDP allocation of the 500 genes. 
}
\label{F:clustering}
\end{center}
\end{figure}

For the DLBCL data, the upper  panel of Figure \ref{F:clustering} displays the estimated posterior density of the  PDP's discount parameter $d$. The estimated posterior probability of the event $[d=0]$ is exactly zero, implying that a non-Dirichlet process clustering mechanism is strongly favored by the data, as suggested earlier by the EDA. The middle   panel of Figure \ref{F:clustering} plots the estimated posterior density of the number of clusters. The a posteriori large number of clusters (for $p=500$ covariates) is suggestive of a PDP model with $d>0$ (i.e.\ a non-Dirichlet process model).
The lower  panel of Figure \ref{F:clustering} summarizes the cluster sizes of the least-squares  allocation \citep{Dahl_2006}. The large number of clusters ($\hat{q}=165$) and the multiplicity of small clusters are very unusual for a Dirichlet process, justifying the use of the more general PDP~model.

\begin{figure}
\begin{center}
\includegraphics[scale=0.3]{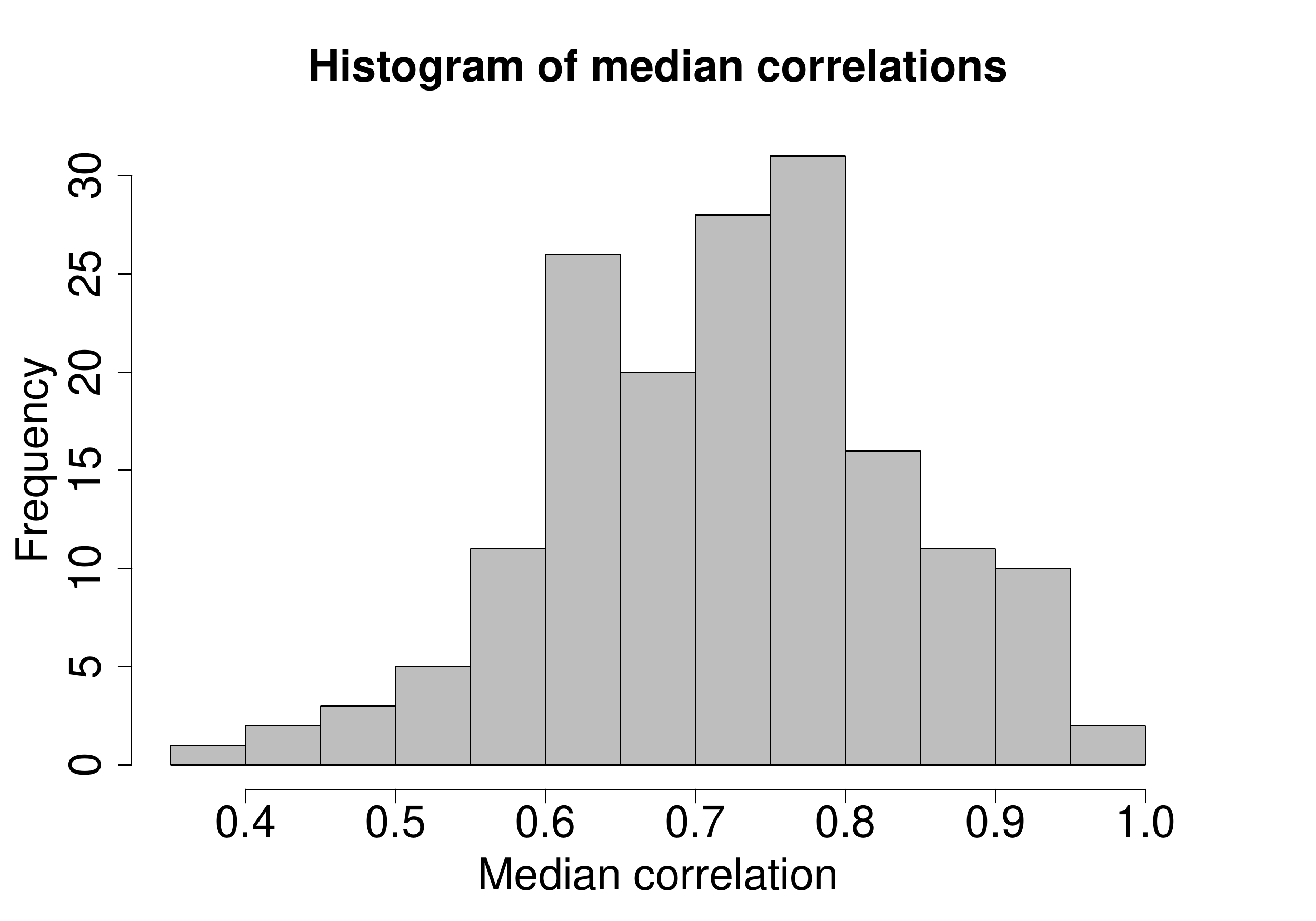}
\includegraphics[scale=0.3]{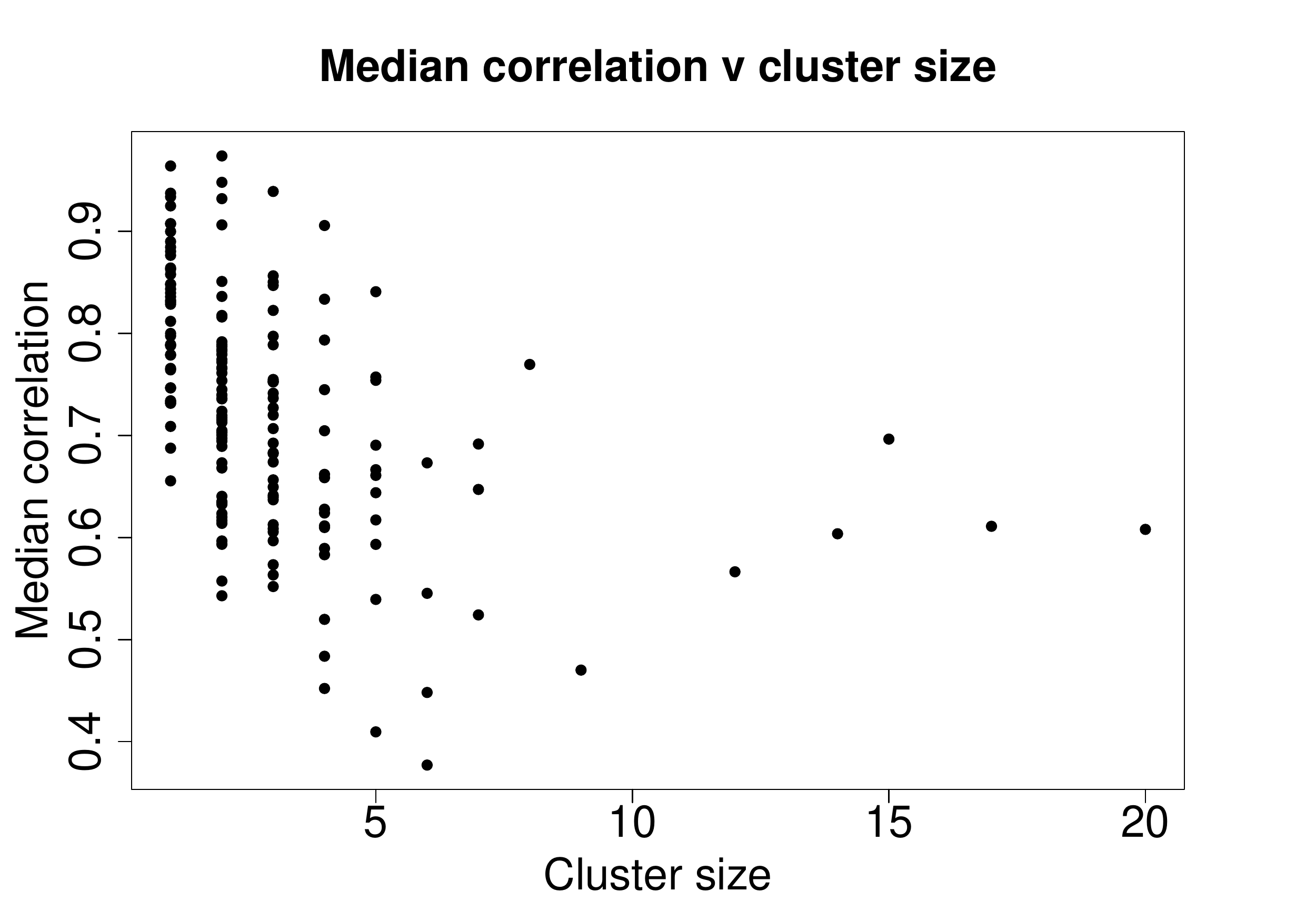}
\caption{For the DLBCL dataset, median pairwise correlations for the $\hat{q}=165$ PDP clusters in the least-squares allocation of Stage 1a.}
\label{F:corr}
\end{center}
\end{figure}

\begin{figure}
\begin{center}
\includegraphics[scale=0.5]{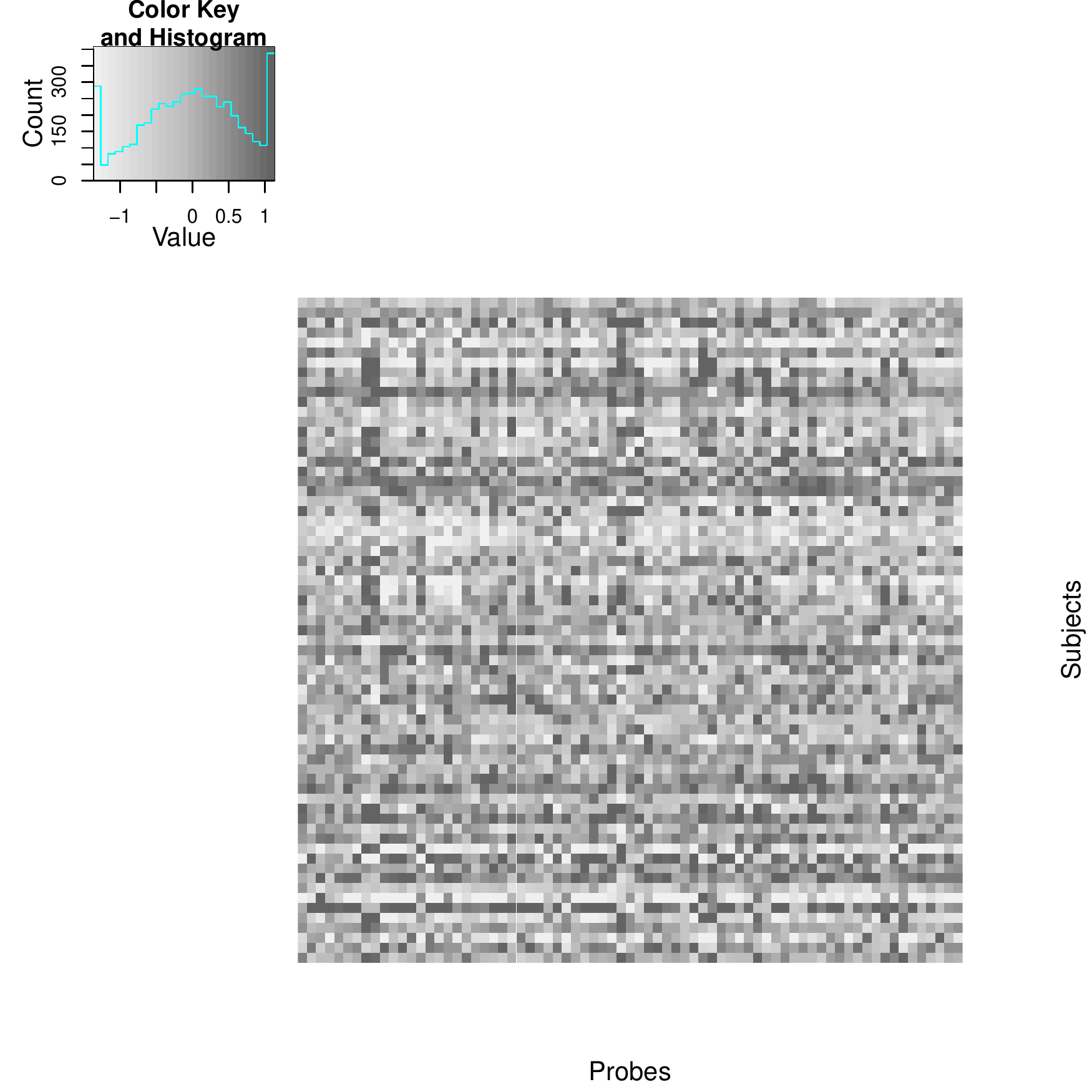}
\includegraphics[scale=0.5]{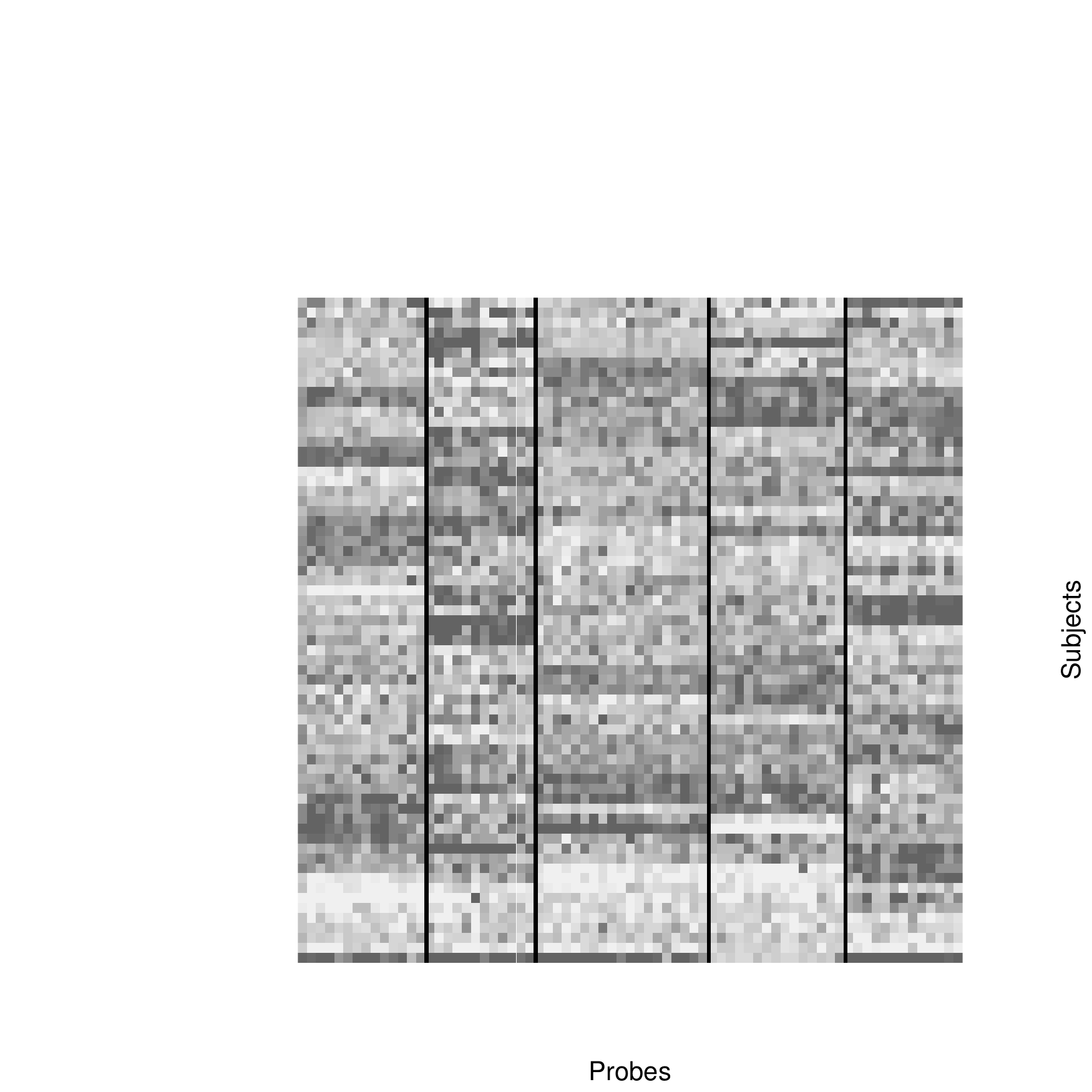}
\caption{Heatmaps of DLBCL covariates that were assigned to latent column clusters with more than 10 members. The panels display the covariates before and after  bidirectional local clustering by VariScan. The vertical lines in the bottom panel mark the covariate-clusters. The color key for both panels is displayed at the top of the plot.}
\label{F:heatmap_cluster}
\end{center}
\end{figure}

The effectiveness of VariScan as a model-based clustering procedure can be shown as follows.
For each of the $\hat{q}=165$ clusters in the least-squares allocation of Stage 1a, we computed the correlations between its member covariates and the latent vector for individuals  with $\hat{z}_{ik}=1$. The cluster-wise median correlations are plotted in Figure \ref{F:corr}. The plots reveal fairly good within-cluster concordance regardless of the cluster size. Figure \ref{F:heatmap_cluster} displays heatmaps for the DLBCL covariates that were allocated to column clusters having more than 10 members.
The panels display the covariates before and after bidirectional clustering of the subjects and probes, with the lower panel of Figure \ref{F:heatmap_cluster} illustrating the within-cluster patterns revealed by VariScan. For each column cluster in the lower panel, the uppermost~rows represent the covariates of any subjects that do not follow the cluster structure and which are better modeled as random noise (i.e., covariates with $\hat{z}_{ik}=0$).

\begin{figure}
\begin{center}
\includegraphics[scale=0.25]{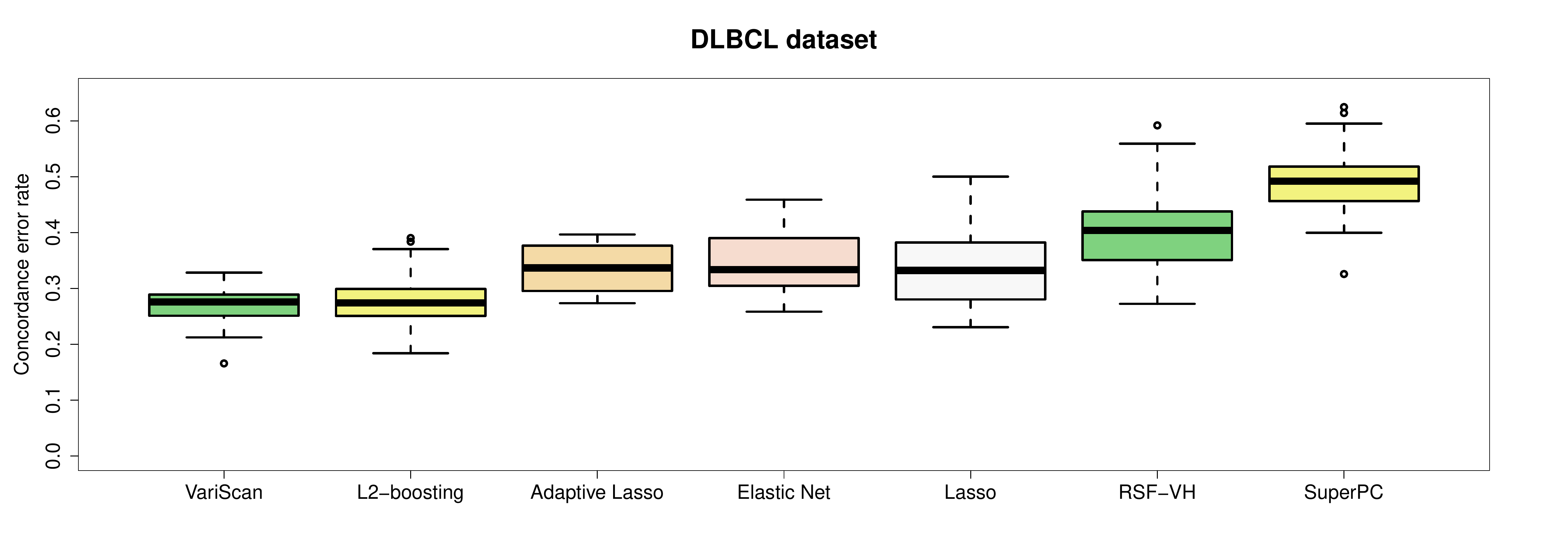}
\includegraphics[scale=0.25]{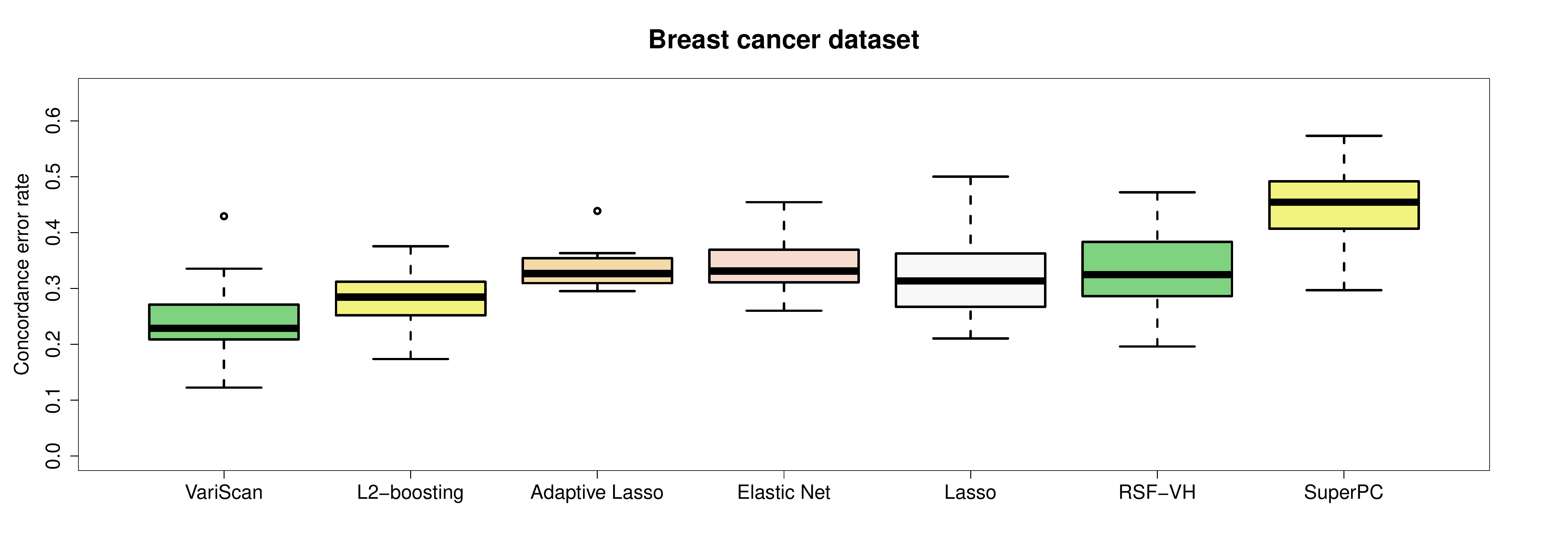}
\caption{Side-by-side boxplots of percentage concordance error rates for the benchmark datasets.}
\label{F:C}
\end{center}
\end{figure}

Comparing the test case predictions with the actual survival times, boxplots of numerical summaries of the concordance error rates for all the methods are presented in Figure \ref{F:C}. 
The success of VariScan appears to be  robust to the different  censoring rates of survival datasets. Although $L_2$-boosting had  comparable error rates for the DLBCL dataset, VariScan had the lowest error rates for both  datasets. Further data analysis results and comparisons are available in Section \ref{S_sup:benchmark data} of the~Appendix.

For subsequent biological interpretations, we selected genes having  high probability of being selected as predictors (with the upper percentile decided by the model size). We then analyzed these genes for their role in cancer progression by cross-referencing with the existing literature. For the breast cancer dataset, our  survey indicated several prominent genes related to breast cancer development and progression, such as  TGF-B2 \citep{pmid17261761}, ABCC3,  which is known to be up-regulated in primary breast cancers, and LAPTM4B, which is related to breast carcinoma relapse with metastasis \citep{pmid20098429}. For the DLBCL dataset, we found several genes related to DLBCL progression, such as the presence of multiple chemokine ligands (CXCL9 and CCL18), interleukin receptors of IL2 and IL5 \citep{pmid16418498},  and BNIP3,  which is down-regulated in DLBCL and is a known marker associated with positive survival \citep{pmid18288132}. A detailed functional/mechanistic analysis of the main set of genes for both datasets is provided in Section \ref{S_sup:benchmark data} of the Appendix.\\

\bigskip

\section{Conclusions}\label{S:conclusion}

Utilizing the sparsity-inducing property of PDPs, VariScan offers an efficient technique for clustering, variable selection, and prediction in high-dimensional regression problems. The covariates are grouped into a smaller number of clusters consisting of covariates with similar across-subject patterns. We theoretically demonstrate  how a PDP allocation can be differentiated from a Dirichlet process allocation in terms of the relative sizes of the latent  clusters. We provide a theoretical explanation for the impressive ability of VariScan to aposteriori detect the true covariate clusters for a general class of models.

In simulations and real data analysis, we show that VariScan makes  highly accurate cluster-related inferences. The technique consistently outperforms established methodologies such as random survival forests, $L_2$-boosting, and supervised principal components, in terms of prediction accuracy. 
In the analyses of benchmark microarray datasets, we identified  several genes having known implications in cancer development and progression, which further engenders our hypothesis.

The VariScan methodology focusses on continuous covariates as a proof of concept, achieving simultaneous clustering, variable selection, and prediction in high-throughput  regression settings  and possessing appealing theoretical and empirical properties. Generalization to count, categorical, and ordinal covariates is possible. It is important to investigate the dependence structures and theoretical properties associated with the more general framework. This will be the focus of our group's future~research.

Due to the intensive nature of the MCMC inference, we performed these analyses in two stages, with cluster detection followed by predictor discovery. We are currently working on implementing VariScan's MCMC procedure in a   parallel computing  framework using  graphical processing units \citep{suchard2010understanding}. We plan to make the software available as an R package for general purpose use in the near future. The single-stage analysis will allow the regression and clustering results to be interrelated, as implied by the VariScan model.  We anticipate being able to  dramatically speed up the calculations by multiple orders of magnitude, which will    allow for single-stage inferences of user-specified datasets  on ordinary desktop and laptop~computers.

\bibliographystyle{plainnat}
\bibliography{EJS}




\end{document}